\documentclass[usenatbib,fleqn]{mn2e}
\usepackage{graphicx,hyperref,amsmath,amssymb,mathrsfs,fixltx2e,listings,caption}
\usepackage[dvipsnames]{xcolor}

\title[AGAMA]
{AGAMA: Action-based galaxy modelling architecture}
\author[E. Vasiliev]{Eugene Vasiliev$^{1,2,3}$\thanks{E-mail: eugvas@lpi.ru}\\
$^1$Rudolf Peierls Centre for Theoretical Physics, Keble road, Oxford, UK, OX1 3NP\\
$^2$Institute of Astronomy, Madingley road, Cambridge, UK, CB3 0HA\\
$^3$Lebedev Physical Institute, Leninsky prospekt 53, Moscow, Russia, 119991}

\begin{document}
\date{Accepted 2018 September 27. Received 2018 September 27; in original form 2018 February 2}
\pagerange{1525--1544}\volume{482}\pubyear{2019}
\setcounter{page}{1525}

\maketitle

\begin{abstract}
\textsc{Agama} is a publicly available software library for a broad range of applications in the field of stellar dynamics.
It provides methods for computing the gravitational potential of arbitrary analytic density profiles or $N$-body models; orbit integration and analysis; transformations between position/velocity and action/angle variables; distribution functions expressed in terms of actions and their moments; iterative construction of self-consistent multicomponent galaxy models. Applications include the inference about the structure of Milky Way or other galaxies from observations of stellar kinematics; preparation of equilibrium initial conditions for $N$-body simulations; analysis of snapshots from simulations.
The library is written in C++, provides a Python interface, and can be coupled to other stellar-dynamical software: \textsc{Amuse}, \textsc{Galpy} and \textsc{Nemo}.
It is hosted at \url{http://github.com/GalacticDynamics-Oxford/Agama}.
\end{abstract}

\begin{keywords}
galaxies: kinematics and dynamics -- methods: numerical
\end{keywords}

%%%%%%%%%%%%%%%%%%%%%%
\section{Introduction}

Galaxy models are vital for understanding their structure and evolution. The rapid increase in quantity and quality of observational data, both for Milky Way and external galaxies, calls for similar advances in modelling techniques. One of the most powerful approaches describes the stars and other mass components by distribution functions (DF) in the space of integrals of motion. For several reasons discussed later in the paper, actions are the most appropriate choice for these integrals of motion. A DF provides a complete description of the system, and various other properties (density, velocity distributions, etc.) can be computed from a DF in a given potential. A flexible representation of the gravitational potential is also a necessity. A dynamically self-consistent model of a stellar system implies certain relations between the potential and the DF; depending on the scientific context, this may or may not be required. 

This paper presents a software framework for galaxy modelling -- \textsc{Agama}. It provides necessary tools for the construction of customized dynamical models described by distribution functions in action space, but many parts of the library, such as general-purpose potential solvers, are applicable to a broader range of problems in stellar dynamics. It is organized into several modules, starting from basic mathematical routines to a complete framework for constructing galaxy models, which are presented in more detail in the following sections:
\begin{itemize}
\item Gravitational potentials, including a few commonly used analytical potential-density pairs  and two versatile potential expansions that can be constructed from an arbitrary density distribution or from an $N$-body model (Section \ref{sec:potential}).
\item Transformation between coordinate/velocity and action/angle variables, in particular, a new implementation of the St\"ackel fudge (Section~\ref{sec:actions}).
\item Several types of DFs expressed in terms of actions (including a new class of disc DF), and associated routines for computing DF moments and creating an $N$-body representation of a DF in a given potential (Section~\ref{sec:df}).
\item The framework for iterative construction of self-consistent galaxy models (Section~\ref{sec:scm}).
\end{itemize}

We illustrate some of the possible applications and compare \textsc{Agama} to other similar software projects in Section~\ref{sec:discussion}.

The \textsc{Agama} library is written primarily in C++ to achieve maximum efficiency, and has a Python interface offering greater flexibility for practical work. It is distributed with many example programs both in C++ and Python, illustrating various aspects of its use (we present a few short Python listings in the paper). It also includes several other stellar-dynamical software projects: an updated version of the Monte Carlo simulation code \textsc{Raga} \citep{Vasiliev2015}, the Fokker--Planck code \textsc{PhaseFlow} \citep{Vasiliev2017}, and parts of the Schwarzschild orbit-superposition code \textsc{Smile} \citep[Vasiliev \& Valluri, in prep.]{Vasiliev2013}.
Extensive documentation \citep{Vasiliev2018} describes the structure and usage of the library, and also contains a more technical description of various methods used in the code (including original implementations of many mathematical tasks such as spline approximation and fitting, sampling from multidimensional distribution functions, etc.) that we do not repeat in this paper.

\textsc{Agama} has interfaces to several other astrophysical frameworks: the potential approximations can be used in \textsc{Amuse} \citep{PortegiesZwart2013}, \textsc{Galpy} \citep{Bovy2015} and \textsc{Nemo} \citep{Teuben1995}, while the routines for action/angle conversion may serve as a more efficient drop-in replacement for the ones in \textsc{Galpy}.
The library is publicly available at \url{http://github.com/GalacticDynamics-Oxford/Agama}.

%%%%%%%%%%%%%%%%%%%%
\section{Potentials}  \label{sec:potential}

At the heart of any galaxy modelling framework lies the gravitational potential.
Some commonly used choices (e.g., Hernquist, Miyamoto--Nagai, or Navarro--Frenk--White models) have analytic expressions for the potential only in spherical or at most axisymmetric cases; some other models (e.g., S\'ersic, exponential disc, or double-power-law halo profiles) require numerical integration or evaluation of expensive special functions. In a general case, one needs to solve the Poisson equation
\begin{align}  \label{eq:Poisson}
\nabla^2 \Phi(\boldsymbol{x}) = 4\pi\,G\,\rho(\boldsymbol{x})
\end{align}
to obtain the potential $\Phi$ of a given density profile $\rho$. \textsc{Agama} provides a few standard potential--density pairs and two general-purpose potential solvers suitable for a wide range of applications.

%%%%%%%%%%%
\subsection{Spherical-harmonic expansion}  \label{sec:potential_sphharm}

Mathematically, the potential can be expressed directly as a three-dimensional integral of the product of the Green's function and the density:
\begin{align}  \label{eq:Poisson3d}
\Phi(\boldsymbol{x}) = -G \iiint d^3x' \, \rho(\boldsymbol{x}') \times \frac{1}{|\boldsymbol{x}-\boldsymbol{x}'|} .
\end{align}
This is computationally challenging both because of the need to compute a triple integral over infinite domain for each point, and because the Green's function is singular. Nevertheless, this approach has been used in the literature \citep[e.g.,][]{Fragkoudi2015}.

A commonly used alternative is to replace two of the three integrals by sums over a certain set of basis functions, such as spherical harmonics \citep[Chapter 2.4]{BinneyTremaine}. Namely, we write the density and potential in spherical coordinates as
\begin{align}  \label{eq:DensitySphHarm}
\{\rho,\Phi\}(r,\theta,\phi) = \sum_{l=0}^\infty \sum_{m=-l}^{l} \{\rho,\Phi\}_{lm}(r)\, Y_l^m(\theta,\phi),
\end{align}
where $Y_l^m$ are the real-valued spherical harmonics (products of associated Legendre functions in $\cos\theta$ and trigonometric functions in $m\phi$). Thanks to the orthogonality of the spherical-harmonic basis, the coefficients of this expansion at each radius $\rho_{lm}$ can be expressed through integrals of the density over two angular variables:
\begin{align}  \label{eq:DensitySphHarmCoefs}
\rho_{lm}(r) = \int_0^{\pi} d\theta \int_0^{2\pi} d\phi\, \rho(r,\theta,\phi) \, Y_l^{m}(\theta,\phi) \, A_{lm},
\end{align}
where $A_{lm}$ are normalization constants. Finally, the coefficients of potential expansion are given by one-dimensional integrals:
\begin{align}  \label{eq:PoissonSphHarm}
\Phi_{lm}(r) = -\frac{4\pi G}{2l+1}\, \bigg[\, r^{-1-l} & \int_0^r dr'\, \rho_{lm}(r')\, r'^{\,l+2}\;+ \\ r^l & \int_r^\infty dr'\, \rho_{lm}(r')\, r'^{\,1-l}  \bigg]. \nonumber
\end{align}

For this approach to be practical, one needs to restrict the range of summation over $l,m$ in the above formulae to some maximum values $0\le l \le l_\mathrm{max}$, $|m| \le \mathrm{min}(l, m_\mathrm{max})$; $m_\mathrm{max}$ may be smaller than $l_\mathrm{max}$ or even zero for axisymmetric systems.
For systems that are not too far from spherical (e.g., elliptical galaxies), $l_\mathrm{max}$ could be as small as 4--8, and in many cases even two terms (monopole and quadrupole) are sufficient. On the other hand, for strongly flattened systems one needs to use many terms in the expansion (for instance, $l_\mathrm{max}=36$ in \citealt{Holley2005}), which is quite demanding computationally.

An elegant modification of this approach was introduced by \citet{KuijkenDubinski1995} for axisymmetric disky systems with a separable density profile: $\rho(R,z) = \rho_R(R)\rho_z(z)$. The key point is that the potential may be split into two parts: one given by an analytic expression involving $\rho_R$ and the second antiderivative of $\rho_z(z)$, and the remainder corresponding to a residual density profile. The latter is not strongly confined to the plane $z=0$ and is efficiently represented by a moderate number of spherical harmonic coefficients. The implementation by W.Dehnen is known as \textsc{GalPot} and has been used in many studies, starting from \citet{DehnenBinney1998}; a nearly equivalent implementation is included in \textsc{Agama}. On the other hand, it is restricted to double-exponential axisymmetric disc profiles, and even though the method can be extended to more general density models, the constraints of separability and axisymmetry cannot be lifted.

%%%%%%%%%%%
\subsection{Azimuthal-harminic expansion}  \label{sec:potential_azimuthal}

Another less well-known possibility is conceptually in between the two described approaches. Namely, the density and potential are both expanded in Fourier harmonics in the azimuthal angle $\phi$, but the expressions relating each harmonic coefficient of potential to its corresponding density coefficient are given by two-dimensional integrals:
\begin{align}  \label{eq:PoissonAziHarm}
\{\rho,\Phi\}(R,z,\phi) &= \sum_{m=0}^\infty \{\rho,\Phi\}_m(R,z)\, \exp(im\phi),
\end{align}
\begin{align*}
\Phi_m(R,z) &= -G\int_{-\infty}^\infty dz' \int_0^\infty dR'\, \rho_m(R',z')\, \Xi_m(R,z,R',z'),
\end{align*}
where $\Xi_m$ is the Green's function in cylindrical coordinates, which has an analytic expression in terms of the Legendre function of the second kind $Q_{m-1/2}$ \citep{CohlTohline1999}. In the axisymmetric case, only $m=0$ term is needed, and in general the above sum may be truncated at a rather moderate $m_\mathrm{max} \lesssim 10$. This potential solver has been introduced in the context of Schwarzschild modelling \citep{VasilievAth2015} and upgraded for the current implementation.

%%%%%%%%%%%
\subsection{Comparison of the two approaches}

In both spherical-harmonic (\ref{eq:PoissonSphHarm}) and azimuthal-harmonic (\ref{eq:PoissonAziHarm}) cases, the potential coefficients $\Phi_{lm}(r)$ and $\Phi_m(R,z)$ are pre-computed at the nodes of a 1d grid in spherical radius $r$ or a 2d grid in the meridional plane, using a moderate number of points (few tens per dimension). The global potential approximation is then constructed by creating interpolating splines in suitably scaled coordinates, which allows one to compute the potential and its two derivatives at any point in space in a very efficient way. In the present implementation, we additionally pre-compute the derivatives of the potential at the grid points, which allows the use of a high-accuracy quintic spline interpolation, introduced in \citet{DehnenBinney1998}. Moreover, following the latter paper, the spherical-harmonic expansion with 1d radial spline interpolation is actually converted to a 2d spline in the $r,\theta$ plane for each value of $m$, which is more efficient than evaluation of Legendre polynomials if the number of harmonics $l_\mathrm{max} > 2$. Note that the difference with the azimuthal-harmonic expansion is in that in the latter case, the 2d spline is constructed for a rectangular grid in $R,z$ coordinates (see also Figure~\ref{fig:density_grids} in Section~\ref{sec:scm_present} for an illustration). The computational effort needed to evaluate the potential and/or its derivatives is thus similar for both methods and comparable to that of \textsc{GalPot}. However, the potentials in \textsc{Agama} can be constructed for any geometry (axisymmetric, triaxial or even more general), while the \textsc{GalPot} approach is limited to axisymmetry.

The two approaches based on harmonic expansion are complementary to each other in several aspects. While the potential evaluation through 2d integration is more accurate than the spherical-harmonic expansion in the case of a highly flattened density distribution, it is also much more computationally intensive to construct. Moreover, it is only accurate enough for density profiles with a finite value at origin and a steep decline at large radii, since the grid in the meridional plane may only cover a finite volume, and the potential outside the grid is extrapolated using a low-degree multipole expansion with a zero Laplacian (hence the extrapolated density is zero). On the other hand, the radial grid in the spherical-harmonic potential approximation covers a wide range of radii (with equally spaced nodes in $\log r$), and the extrapolation to small and large radii is able to represent power-law density profiles outside the grid fairly accurately. Thus the latter approximation is more suitable for spheroidal galaxy components, such as the bulge and the halo, while the former is adequate for the disc component. One may get the best of both worlds by combining the two potential expansions, each one representing one or more density components, into a composite potential.

%%%%%%%%%%%
\subsection{Smooth approximations to $N$-body potentials}

Both potential expansions may be computed either from an analytic density profile, or from a set of $N$ point masses, thus approximating the potential of an $N$-body system by a smooth one. In fact, as shown in \citet{Vasiliev2013} and \citet{VasilievAth2015}, in test cases when the particle positions are sampled from a known analytic density model, the smooth potential expansion more closely matches the corresponding analytic profile than a potential computed directly from the $N$-body snapshot, e.g., using a tree-code method, while also being much faster to evaluate. Of course, in this case the error in potential is dominated by discreteness noise rather than interpolation error, but typical $N$-body potential solvers suffer even more from this noise. Thus the potential expansion approach is well suited for analyzing the properties of orbits in a `frozen' potential of an $N$-body system, and could be easily extended to represent a time-dependent potential whose coefficients are interpolated in time, providing a more flexible alternative to parametrized analytic models \citep[used, e.g., in][]{Muzzio2005,MachadoManos2016} or tree-code potentials \citep[e.g.,][]{Valluri2010}. The computation of a smooth potential from discrete samples is at the core of the Monte Carlo simulation code \textsc{Raga} \citep{Vasiliev2015}, which is also included in the framework.

%%%%%%%%%%%
\subsection{Miscellanea}

For completeness, we mention another approach that replaces all three integrals in (\ref{eq:Poisson3d}) with sums over members of a basis set, which typically consists of products of spherical-harmonic functions and certain families of orthogonal polynomials in scaled radial coordinate \citep[e.g.][]{Zhao1996,Lilley2018}. This approach is commonly associated with the `self-consistent field' method \citep{HernquistOstriker1992,Meiron2014} and is similar to the spline-interpolated spherical-harmonic expansion, but is less flexible and more computationally demanding. To compute the potential of a basis-set expansion, one needs to sum over all basis functions, while for the case of spline interpolation only the values at two adjacent grid nodes are required. \citet{Vasiliev2013} compared both approaches and found the spline interpolation to be generally faster and more accurate; therefore, the basis-set expansion is no longer included in the library. %[Kalnajs 1976, ApJ, 205, 745]

%%%%%%%%%%%%%%%%%
\begin{lstlisting}[
caption={Gravitational potentials constructed from smooth density profiles or from an $N$-body snapshot. \protect\\
We create a three-component model of a galaxy with a Sersic bulge (flattened in $z$ direction), an exponential disc, and a Navarro--Frenk--White halo. %The first one is represented as a multipole expansion (chosen implicitly because the input density is of a spheroid family), the second one -- using the \textsc{GalPot} approach (a combination of an analytic part of the disc and a multipole expansion of the residual), and the third one is a simple spherical analytic profile.
We also sample $10^5$ particles from the bulge density profile and use this $N$-body snapshot to construct another multipole potential.%, manually assigning the level of symmetry (which determines the choice of non-trivial angular harmonics). 
\protect\\
Then we illustrate the properties of the composite potential by plotting the rotation curve $v_\circ(R) = \sqrt{R\: \partial\Phi/\partial R}$ for each component separately and for the total potential, and the density of each component in the equatorial plane.\protect\\
As could be seen by running this script, the density profile of the smooth potential constructed from the $N$-body snapshot differs from the analytic one by $\lesssim 1\%$.
}, label=code:potential, float=t]
import agama, numpy, matplotlib.pyplot as plt

# create three components of a composite galaxy potential
pot_bulge = agama.Potential(type='Sersic', mass=1,
    scaleRadius=1, sersicIndex=4, axisRatioZ=0.6)
pot_disc = agama.Potential(type='Disk', mass=4,
    scaleRadius=3, scaleHeight=0.5)
pot_halo = agama.Potential(type='NFW', mass=25,
    scaleRadius=10)
pot = agama.Potential(pot_bulge, pot_disc, pot_halo)

# represent the density profile as a collection of particles
snap = pot_bulge.sample(100000)

# create a potential from this N-body snapshot
pot_nbody = agama.Potential(type='Multipole',
    particles=snap, symmetry='Axisymmetric')

# choose the grid in radius to plot the profiles
r=numpy.linspace(0.0, 25.0, 250)
xyz=numpy.column_stack((r, r*0, r*0))

# circular velocity as a function of radius: total...
vcirc_total = numpy.sqrt(-r * pot.force(xyz)[:,0])
plt.plot(r, vcirc_total, label='Total')
# ...and for each potential component separately
for p in pot:
    plt.plot(r, numpy.sqrt(-r * p.force(xyz)[:,0]),
    label=p.name() )
plt.legend(loc='lower right')
plt.show()

# density in the equatorial plane for each component
for p in pot:
    plt.plot(r, p.density(xyz), label=p.name() )
plt.plot(r, pot_nbody.density(xyz), label='from Nbody')
plt.yscale('log')
plt.legend()
plt.show()
\end{lstlisting}
%%%%%%%%%%%%%%%%%

Listing~\ref{code:potential} illustrates the construction and use of various potential types, and the creation of a smooth potential approximation from an $N$-body snapshot (in this example, the snapshot itself was generated from a smooth density profile; in realistic applications it would be taken from an $N$-body simulation).

Given a potential, it is straightforward to numerically integrate orbits; we use a modified version of the $8^\mathrm{th}$ order Runge--Kutta integrator \textsc{dop853} \citep{DOP853}. All potentials in the library provide up to two derivatives, which may be used for computing the largest Lyapunov exponent (the corresponding routine is included in \textsc{Agama}) or other variational chaos indicators \citep[e.g.,][]{Carpintero2014}.

%%%%%%%%%%%%%%%%%%%%%%%%%%%%%%%%
\section{Action/angle variables}  \label{sec:actions}

\textsc{Agama} deals with models of stellar systems expressed in terms of action/angle variables $\boldsymbol{J},\,\boldsymbol{\theta}$ (see \citealt{BinneyTremaine}, Section~3.5, for the definition), and provides several methods for conversion between position/velocity and action/angle variables. The most practically important one is the axisymmetric St\"ackel fudge \citep{Binney2012}, for which our implementation is more efficient and accurate than other existing codes, thanks to an improved method for choosing the focal distance. In this section we review the action/angle formalism and present in detail the approach used in the library.

%%%%%%%%%%%
\subsection{General properties}  \label{sec:actions_general}

Strictly speaking, action/angle variables are only well-defined when the motion is integrable (multiperiodic); the meaning of actions depends on the orbit type, and in a spherical or an axisymmetric potential, the most convenient choice is the triplet $\{J_r, J_z, J_\phi\}$. The first two (radial and vertical actions) describe the extent of oscillations in spherical radius and vertical dimension (for a nearly-circular orbit in the equatorial plane they decouple, otherwise the motion is qualitatively similar but more complex), while the third (azimuthal action) is the conserved component $L_z$ of the angular momentum. In the spherical case $J_z+|J_\phi|$ is the total angular momentum $L$. 

Actions are just another set of integrals of motion, so they can be expressed in terms of more familiar integrals such as energy $E$, angular momentum $L$ or $L_z$, and the non-classical third integral $I_3$ (assuming that it exists). They have certain advantages over other choices of integrals: 
\begin{itemize}
\item Action/angle variables are canonical, i.e., satisfy the Hamilton's equations of motion, which in this case are trivial: $\boldsymbol{J} = \mathrm{const}$, $\dot{\boldsymbol{\theta}} = \mathrm{const} = \boldsymbol{\Omega} \equiv \partial H/\partial \boldsymbol{J}$, where $H$ is the Hamiltonian expressed in terms of actions.
\item The possible range of each action variable is $[0,\infty)$ or $(-\infty,\infty)$, independently of others (unlike, say, $E$ and $L$).
\item The transformation between $\boldsymbol{x}, \boldsymbol{v}$ and $\boldsymbol{J}, \boldsymbol{\theta}$ has unit determinant, which is convenient for dealing with DFs (for instance, the mass of the system is given by $\int f(J)\; d^3J\,d^3\theta$ independently of the potential), see Section~\ref{sec:df}.
\item Actions are adiabatical invariants, conserved under slow changes in the potential. This is convenient for iterative self-consistent modelling (Section~\ref{sec:scm}), and simplifies the treatment of multicomponent systems (e.g., growing a stellar disc in a dark halo preserves the action-based DF of the latter, even though the correspondence between actions and coordinates changes).
\item Action/angle variables are naturally suited to analyze perturbations from equilibrium state \citep[e.g.,][]{Monari2016} and collisional relaxation \citep[e.g.,][]{Fouvry2015}.
\end{itemize}

There exist several methods for conversion between position/velocity and action/angle variables (see \citealt{SandersBinney2016} for a comprehensive review and comparison of approaches). It can be performed in both directions analytically in special cases, such as the Isochrone potential, or using 1d numerical quadratures for an arbitrary St\"ackel potential (in particular, a spherical one); these methods are available in the library. 
For a more interesting practical case of an arbitrary axisymmetric potential, we use the `St\"ackel fudge' \citep{Binney2012}, which is an approximate method for computing actions and angles under the assumption that the motion is integrable and is locally well described by a St\"ackel potential, separable in prolate spheroidal coordinates. \textsc{Agama} contains a fresh implementation of this approach, which is more efficient and accurate than other existing codes (\textsc{tact}, \citealt{SandersBinney2016}, or \textsc{Galpy}, \citealt{Bovy2015}).
The reverse transformation (from action/angles to position/velocity) is provided by the \textsc{TorusMapper} package \citep{BinneyMcMillan2016}, which is included in the \textsc{Agama} library with some modifications (most notably, an improved angle mapping method used by \citealt{BinneyKumar1993}, \citealt{LaaksoKaasalainen2013}).
At the moment the action/angle framework applies only to oblate axisymmetric potentials, although there exist methods suitable for triaxial potentials \citep{SandersBinney2014,SandersBinney2015a}.

%%%%%%%%%%%%%%
\begin{figure}
\includegraphics{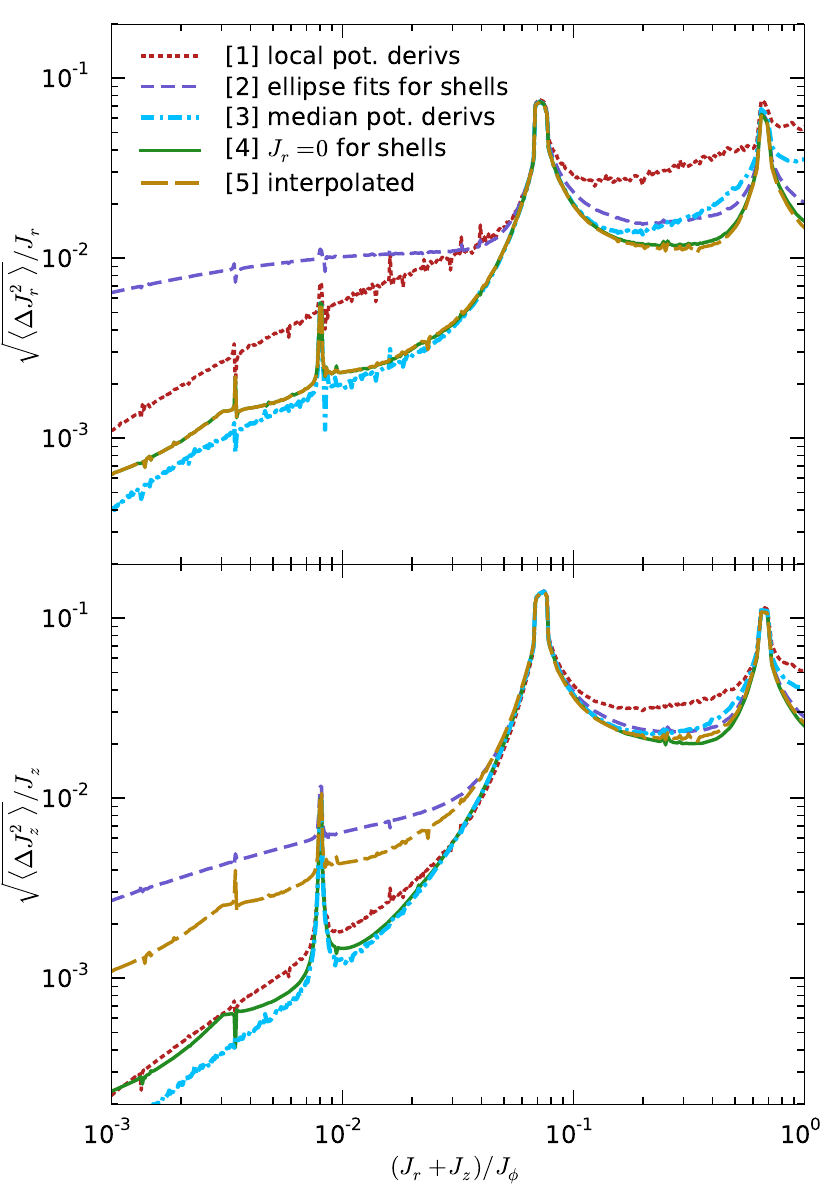}
\caption{Accuracy of the St\"ackel fudge for various methods of computing the focal distance $\Delta$.
Shown is the relative r.m.s. variation of the radial (top panel) and vertical (bottom panel) action for a series of orbits in a realistic galactic potential \citep{Piffl2014}, started at $R=8.3$~kpc, $z=0$, with a total velocity 240~km/s (equal to the local circular speed), split in different proportions between $v_R, v_z=0.8v_R$, and $v_\phi$. These initial conditions are similar to the ones considered in Section~5.2 and plotted in Figure~3 of \citet{SandersBinney2016}, except that we keep the total velocity constant, not $v_\phi$, and plot the relative r.m.s. variation instead of the absolute one. We numerically integrate each orbit for 10 dynamical times and compute the variation of the actions evaluated at 1000 points sampled from the trajectory.\protect\\
Dashed blue curve [2] corresponds to $\Delta$ being interpolated from a pre-initialized 2d grid in $E,L_z$ using ellipse fits to shell orbits, as in \citet{Binney2014}; it matches the `Fudge v2' method in \citet{SandersBinney2016}. Red dotted line [1] corresponds to $\Delta$ being evaluated separately for each input point from the second derivatives of the potential (equation~15 in that paper), it matches the `Fudge v1' method. Cyan dot-dashed line [3] uses a similar approach, but taking the median value of $\Delta$ for the entire trajectory (as used in \textsc{Galpy}). Green solid line [4] takes $\Delta$ from a pre-initialized 2d grid in $E,L_z$, using the condition that $J_r=0$ for shell orbits (equation~\ref{eq:DeltaShell}); this is the new method introduced in \textsc{Agama}. Yellow long-dashed line [5] uses the same $\Delta$, but interpolates actions from a pre-initialized 3d grid in $E,L_z,I_3$.\protect\\
Naturally, the accuracy is best for low-eccentricity orbits (with $J_r,J_z \ll J_\phi$), and worst for resonant orbits (several strong peaks in both plots). For the former, the median value of $\Delta$ for the entire orbit [5] delivers the highest accuracy, but of course requires the orbit to be computed first. Given just a single point, equation~\ref{eq:DeltaShell} produces the most accurate estimate [4], and its interpolated variant [5] is only moderately worse but $\sim10\times$ faster.
} \label{fig:action_accuracy}
\end{figure}

%%%%%%%%%%%%%%
\begin{figure}
\includegraphics{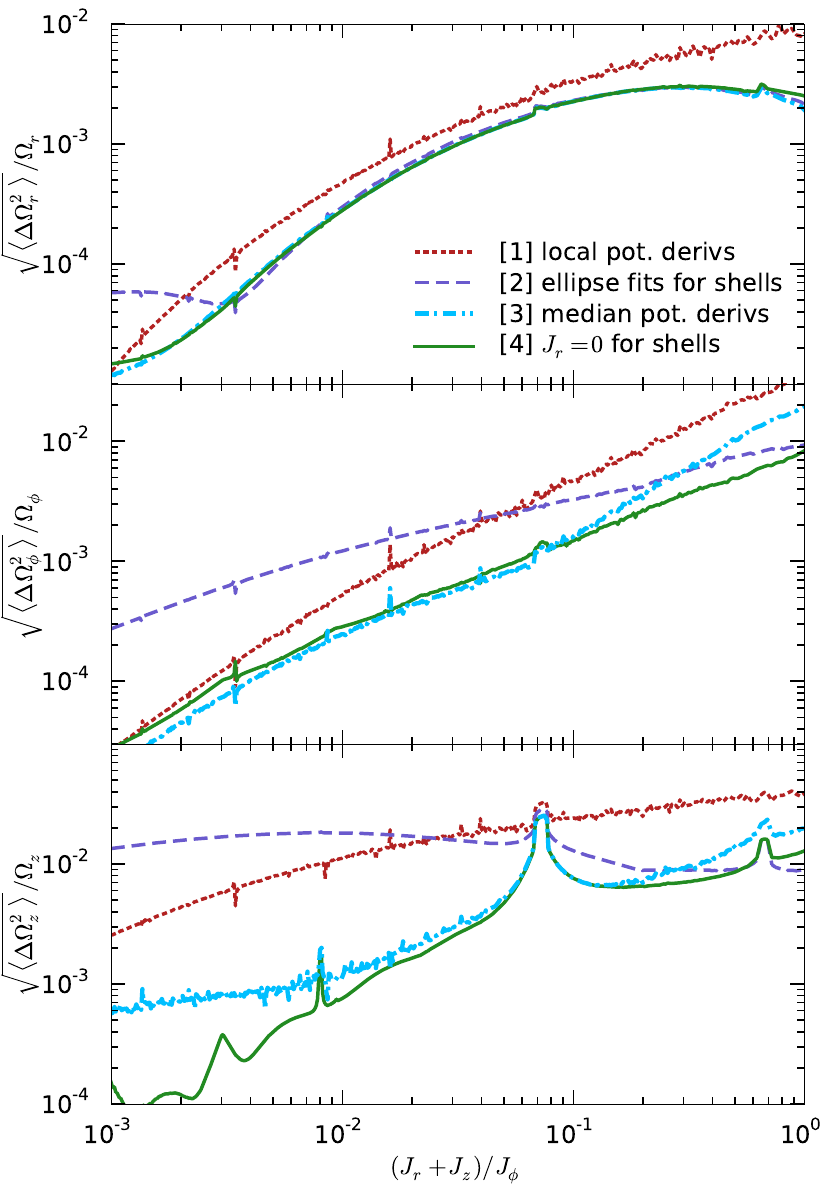}
\caption{Accuracy of frequency determination by the St\"ackel fudge for various choices of focal distance.
This plot is similar to Figure~\ref{fig:action_accuracy} and uses the same orbits and methods (except the interpolated one, in which frequency determination is not implemented), but shows the relative r.m.s. variation of three frequencies; it may be compared to Figure~4 in \citet{SandersBinney2016}.
Again the choice of $\Delta$ in \textsc{Agama} (green solid line [4]) produces the most accurate results.
} \label{fig:frequency_accuracy}
\end{figure}

%%%%%%%%%%%%%%
\begin{figure}
\includegraphics{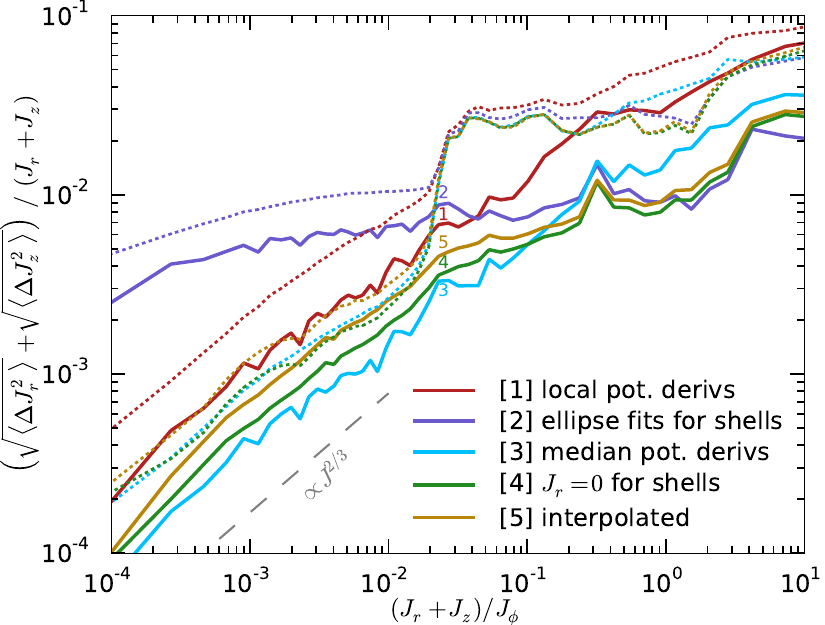}
\caption{Accuracy of the St\"ackel fudge for various choices of focal distance $\Delta$.\protect\\
We take a large sample of orbits started at $R=8.3$~kpc, $z=0$ and randomly directed velocity $v=240$~km/s in the Milky Way potential from \citet{Piffl2014}, covering the range from thin disc to halo stars. For each numerically integrated orbit we compute the relative r.m.s. variation of the sum of radial and vertical actions, and plot the median error (solid lines) and 90\% percentile (dotted lines) for the same five methods as in Figure~\ref{fig:action_accuracy}).
The ranking of the methods turns out to be the same, with the one used in \textsc{Agama} being optimal among the methods that use only a single point on the trajectory, and its interpolated version being on average 50\% less accurate. However, a significant fraction of high-eccentricity orbits turn out to be resonant, with much larger variations of actions (seen as a rapid increase in the 90\% percentile error for $(J_r+J_z)/J_\phi \gtrsim 0.02$). For these orbits the St\"ackel fudge is not accurate regardless of the choice of $\Delta$. \protect\\
We checked that the ranking between methods is similar for other values of energy and choices of potential, although the severity of the accuracy deterioration near resonances depends on the properties of potential: for instance, it is almost negligible for the \texttt{MWPotential2014} used in \citet{Bovy2015}, because the latter does not have a thin gaseous disc, which creates strong gradients in vertical frequency in the potential models used by \citet{Piffl2014,McMillan2017}.
} \label{fig:action_accuracy_overall}
\end{figure}

%%%%%%%%%%%%%%%
\begin{figure*}
\includegraphics{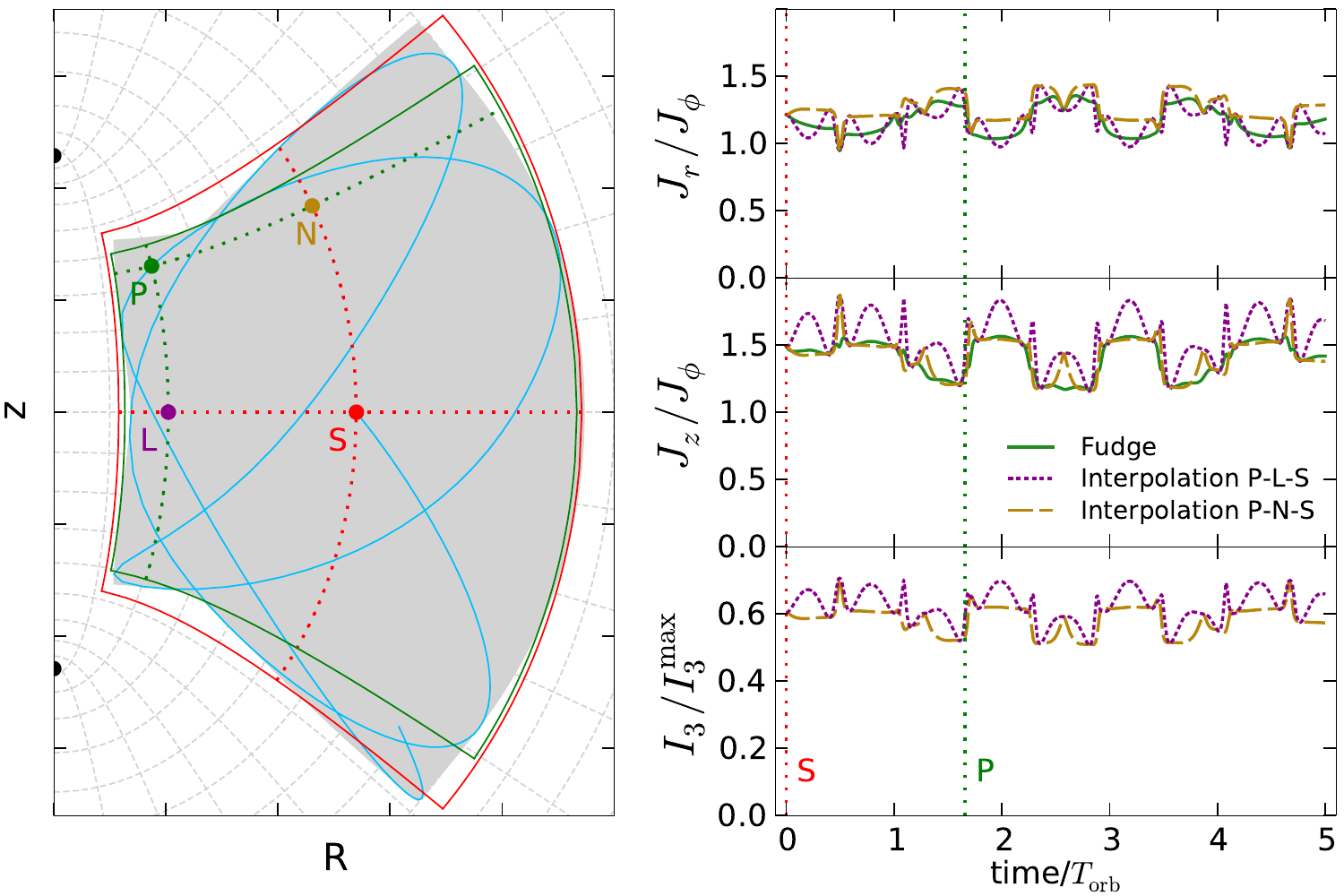}
\caption{Illustration of the St\"ackel fudge and its interpolated variant.
\protect\\
Left panel shows the meridional cross-section of an orbit in a realistic axisymmetric Milky Way potential. The orbit starts at point $S$ with $R=R_\mathrm{shell}(E,L_z), z=0$, and meridional velocity directed at $\sim50^\circ$ to the equatorial plane and being 4.5 larger than the azimuthal velocity (hence the orbit is fairly eccentric and the inaccuracies in action computation are greater than for the majority of stellar orbits). Blue curve shows the initial part of the trajectory, and the entire region spanned by the orbit is shaded in grey. 
Dotted grey lines show the auxiliary prolate spheroidal coordinate system, and black dots on the $z$ axis show its focal points located at $z=\pm \Delta$. \protect\\
To compute actions for any point $P$, we first determine the extent of oscillation in the radial ($\lambda$) and vertical ($\nu$) directions, and then integrate the canonically conjugate momenta $p_\lambda, p_\nu$ along the lines of constant $\lambda$ and $\nu$, shown by green dots. If the potential were of a St\"ackel form, the orbit would align with the prolate spheroidal coordinates, and its extent would be the same for all points on the trajectory; in reality the match is not perfect, and consequently the determined orbit boundaries vary between points (red and green boxes for the points $S$ and $P$, respectively). \protect\\
Alternatively, we `translate' the point $P$ back to the initial point $S$ and determine the value of $I_3$ which is related to the initial velocity direction; the actions are then interpolated from a pre-computed table, which was filled using integration contours originating from $S$ (dotted red curves). There are various possible ways of computing $I_3$: by following the line of $\lambda=\mathrm{const}$ from $P$ to the point $L$ and then the line $\nu=\mathrm{const}$ from $L$ to $S$ (equation~\ref{eq:deltaLambda}, resulting in dotted purple curves), or in the opposite order, from $P$ to $N$ and then to $S$ (equation~\ref{eq:deltaNu}, dashed yellow curves). \protect\\
Right panels show the variation of actions and $I_3$ along this orbit; the two points $S$ and $P$ are marked by vertical dotted lines. The two methods for estimating $I_3$ produce different results, as shown in the bottom sub-panel. The corresponding interpolated actions are shown in the same colour and line style in the other two sub-panels, and their variation mirrors that of $I_3$. The actions computed by the non-interpolated St\"ackel fudge are shown in green curves and exhibit smaller variations, because the errors in determining the orbit boundaries are somewhat compensated by using different integration paths for each point. Hence the interpolated St\"ackel fudge is generally less accurate not because of interpolation itself, but because of larger fluctuations in the approximate third integral. 
Moreover, the fluctuations in actions are anticorrelated, therefore the value of a typical distribution function which depends on a linear combination of $J_r$ and $J_z$, such as equation~\ref{eq:DFhalo}, is computed more accurately than either of the two actions.
} \label{fig:staeckel_orbit}
\end{figure*}
%%%%%%%%%%%%

%%%%%%%%%%%
\subsection{St\"ackel fudge and the role of focal distance}  \label{sec:actions_staeckel}

We now recall the key ingredients of the St\"ackel fudge method, which is the main workhorse in \textsc{Agama}.

Actions and angles can be computed exactly for a general potential separable in a confocal ellipsoidal coordinate system \citep[e.g.,][]{deZeeuw1985}. We specialize to the case of oblate axisymmetric potentials, as being more relevant for flattened disc galaxies. It is possible to extend the formalism to prolate potentials, but an additional complication will be the existence of two types of tube orbits (inner and outer long-axis tubes), hence it is not implemented at the moment.
The potential is expressed in a prolate spheroidal coordinate system defined by the focal distance $\Delta$. The transformation between the cylindrical coordinates $R,z$ and spheroidal coordinates $\lambda,\nu$ is given, e.g., by equation~8 in \citet{Sanders2012}. A St\"ackel potential has a form 
\begin{align}  \label{eq:StaeckelPot}
\Phi(\lambda,\nu) = -\frac{f_\lambda(\lambda)-f_\nu(\nu)}{\lambda-\nu} ,
\end{align}
in other words, instead of being an arbitrary function of two coordinates, it is defined by two one-dimensional functions. An orbit in such a potential respects three integrals of motion -- energy $E$, $z$-component of angular momentum $L_z$, and a non-classical third integral%
\footnote{There are various definitions of $I_3$ in the literature, but they are equivalent up to a shift or multiplication by a constant that depends on $E$, $L_z$ and $\Delta$.}
\begin{align}  \label{eq:I3}
I_3 \equiv \frac{z^2\,v_\phi^2 + (zv_R-Rv_z)^2 + \Delta^2\,v_z^2}{2} +
\frac{\lambda f_\nu(\nu) - \nu f_\lambda(\lambda)}{\lambda-\nu} ,
\end{align}
which ranges from $I_3=0$ for an equatorial-plane orbit with $J_z=0$ to $I_3^\mathrm{max}\equiv (R_\mathrm{shell}^2+\Delta^2)\,v_z^2/2$ for a thin (`shell') orbit that crosses the plane $z=0$ at a single radius $R_\mathrm{shell}(E,L_z)$ with vertical velocity $v_z$ and has $J_r=0$. Given the three integrals, one can numerically solve the equations defining the turn-around points (min/max values of $\lambda,\nu$), and then compute the actions, angles and frequencies by 1d numerical integration.

The essence of the St\"ackel fudge is to use the same expressions, but substitute the real potential instead, which is not of a St\"ackel form (without explicitly constructing an approximating St\"ackel potential, as in \citealt{Sanders2012}). Actions computed in this way are approximate, i.e., not conserved along a numerically integrated orbit. The accuracy of approximation depends on the only free parameter in the method -- the focal distance $\Delta$, which can be different for each point. We now discuss various methods for choosing $\Delta$ and introduce a new one, which appears to be close to optimal.

In a St\"ackel potential, $\Delta$ can be found from the combination of second derivatives of the potential (equation~15 in \citealt{SandersBinney2016}), hence one possible choice is to use this equation in the real potential at each input point (dubbed `Fudge v1' in that paper, and shown by red curves labelled [1] in Figures~\ref{fig:action_accuracy}--\ref{fig:action_accuracy_overall}). Another possibility is to relate $\Delta$ to the quantities conserved for each orbit, namely the two classical integrals of motion $E$ and $L_z$. Since the focal distance is more important for high-inclination orbits and irrelevant for in-plane orbits, \citet{Binney2014} suggested to estimate it for the extreme `shell' orbit, constructed numerically (for each pair $E,L_z$, locating the radius $R_\mathrm{shell}$ such that an orbit launched vertically from the equatorial plane returns back to the same radius). Then $\Delta$ can be computed as the focal distance of an ellipse that best matches the shell orbit.
A 2d interpolation grid for $\Delta(E,L_z)$ is pre-initialized for the given potential and used to find the focal distance for any input point; this method is dubbed `Fudge v2' in \citet{SandersBinney2016} and shown by blue lines [2].

We find that a more suitable choice of $\Delta$ is obtained if one demands that $J_r=0$ for the same shell orbit. Namely, we write the momentum $p_\lambda$, canonically conjugate to the radial coordinate $\lambda$, as a function of $\lambda$, $\Phi(\lambda,\nu=0)$, $E$, $L_z$ and $I_3$ (the latter depends on $\Delta$). For a shell orbit, the range of oscillation in $\lambda$, determined by the condition $p_\lambda^2\ge 0$, collapses to a single point $\lambda_\mathrm{shell}=R_\mathrm{shell}^2+\Delta^2$; hence $\partial p_\lambda^2/\partial\lambda=0$, which translates to
\begin{equation}  \label{eq:DeltaShell}
\Delta^2 = R_\mathrm{shell}^2\; \left.
\frac{2 \big[E - \Phi \big] - R \frac{\partial\Phi}{\partial R} }
{R \frac{\partial\Phi}{\partial R} - \frac{L_z^2}{R^2} } \right|_{R = R_\mathrm{shell},z=0}
\end{equation}

In some cases, this expression produces negative $\Delta^2$, which are replaced by zeros (essentially rendering the approximate St\"ackel potential spherical). We pre-initialize a 2d interpolation table for $\Delta(E,L_z)$, which is a one-time effort for the given potential, requiring a few CPU seconds.
Figures~\ref{fig:action_accuracy} to \ref{fig:action_accuracy_overall} compare the accuracy of St\"ackel fudge for various choices of $\Delta$, demonstrating that equation~\ref{eq:DeltaShell}, shown by the green curve [4], delivers better results than either of the two methods used in \citet{SandersBinney2016}.

In principle, a further improvement (cyan curve [5]) can sometimes be achieved by taking a median value of $\Delta$ from potential derivatives along a numerically computed trajectory (this method is offered in \textsc{Galpy}); however, the overhead of orbit integration makes it impractical for calculating the actions at a single point. \textsc{Galpy} does not contain any method to retrieve a suitable $\Delta$ from an interpolation table, hence a single value has to be used for the entire system, substantially deteriorating the accuracy of action computation.

We checked that for a given $\Delta$, the numerical procedures used in \textsc{Tact}, \textsc{Galpy} and \textsc{Agama} produce nearly identical results; therefore, the improvement in accuracy in \textsc{Agama} arises from a more judicious assignment of $\Delta$ that varies across the phase space.
The cost of action computation for each input point comes from numerical root-finding and integration routines, which require $\sim 50$ potential evaluations; this is $1.5-2\times$ fewer in \textsc{Agama} than in other existing implementations, thanks to various mathematical improvements.

%%%%%%%%%%%
\subsection{Interpolation}  \label{sec:actions_interpolation}

One could achieve considerable savings by pre-computing a 3d interpolation table for actions $J_r,J_z$ as functions of three integrals of motion (one of them being approximate), as part of the initialization procedure, approximately doubling its computational cost. For an orbit started at $R=R_\mathrm{shell}(E,L_z)$, $z=0$ and meridional velocity $v_\mathrm{mer}=\sqrt{2\big[E-\Phi(R_\mathrm{shell})\big] - (L_z/R_\mathrm{shell})^2}$ directed at an angle $\psi$ to the equatorial plane, the value of the third integral (\ref{eq:I3}) is $I_3=(R_\mathrm{shell}^2+\Delta^2)\,v_\mathrm{mer}^2\,\sin^2\psi$, varying from 0 for an in-plane orbit ($\psi=0$) to $I_3^\mathrm{max}(E,L_z)$ for a shell orbit ($\psi=\pi/2$). For each combination of $E$ and $L_z$ used in constructing the interpolator for $\Delta$, we pre-compute $J_r$ and $J_z$ on a grid in $\psi$ (or equivalently $I_3/I_3^\mathrm{max}$) and construct 3d cubic spline interpolators in $E$, $L_z/L_\mathrm{circ}(E)$ and $\psi$ for both actions. Then for an arbitrary point $\{\boldsymbol{x},\boldsymbol{v}\}$ we need to determine the value of $I_3$ to retrieve the actions from the interpolation table. If the potential were of the St\"ackel form (\ref{eq:StaeckelPot}), the third integral would be given by Equation~\ref{eq:I3}. One could compute it without explicitly specifying the functions $f_\lambda(\lambda), f_\nu(\nu)$, by replacing the second term in equation (\ref{eq:I3}) by either of the two expressions:
\begin{subequations}
\begin{align}
f_\lambda(\lambda) + \lambda \Phi(\lambda,\nu) &=
\lambda\;\big[ \Phi(\lambda,\nu) - \Phi(\lambda,0) \big],  \label{eq:deltaLambda}\\
f_\nu(\nu) + \nu \Phi(\lambda,\nu) &=
\nu\;\big[ \Phi(\lambda,\nu) - \Phi(\lambda_\mathrm{shell},\nu) \big]  \label{eq:deltaNu}\\
 &+ \lambda_\mathrm{shell}\;\big[ \Phi(\lambda_\mathrm{shell},\nu) - \Phi(\lambda_\mathrm{shell},0) \big],  \nonumber
\end{align}
\end{subequations}
where the right-hand sides only contain the values of potential at certain points. For a non-St\"ackel potential these expressions are not equivalent and yield different numerical values for $I_3$, which furthermore vary along the orbit (Figure~\ref{fig:staeckel_orbit}, bottom right panel). The former one is related to the radial energy $E_r$ \citep[Eq.18]{Binney2012}, and the latter -- to the vertical energy $E_z$ \citep[Eq.2]{Bovy2015}. The latter paper suggested to use both approximate values of $I_3$ to compute interpolated $J_r$ and $J_z$, correspondingly, whereas Binney's approach only uses the latter one for both actions. We also do the same since we find that (\ref{eq:deltaNu}) is typically better conserved along orbits.

Interestingly, the interpolated actions obtained from either expression for $I_3$ have larger variations than the actions computed by the St\"ackel fudge: the fluctuations in $I_3$ are largely compensated by using different contours ($\lambda=\mathrm{const}$ and $\nu=\mathrm{const}$) for each point $\lambda,\nu$ along the orbit, as shown in the left panel of Figure~\ref{fig:staeckel_orbit}. Hence the lower accuracy of the interpolated actions arises not from the interpolation itself, but from the approximate nature of the third integral. Because the variation of $I_3$ (relative to the maximum value $I_3^\mathrm{max}$) is typically larger than the variations of actions computed by the (non-interpolated) St\"ackel fudge, this suggests that actions have further advantages compared to the `vanilla' third integral, as used, e.g., in \citet{Famaey2002}, \citet{Bienayme2015}. 

Besides larger fluctuations, a more serious problem is that the interpolated actions have small but non-negligible systematic bias (compare the solid and dashed or dotted curves in the top and middle-right panels of Figure~\ref{fig:staeckel_orbit}). This bias depends on the potential and exists for any choice of the approximation for $I_3$, and its magnitude and even sign vary across the action space, being more severe for high-eccentricity orbits. This precludes the use of interpolated actions in applications that require an unbiased comparison between different potentials, such as the maximum-likelihood determination of potential parameters using an action-based DF of kinematic tracers (Section~\ref{sec:DFapplications}). However, in other contexts a moderate deterioration of accuracy in the interpolated actions is acceptable and leads to a $10\times$ increase in computational efficiency, therefore the library offers a choice between using interpolated or non-interpolated actions (in both cases it still constructs a 2d interpolation table for $\Delta(E,L_z)$ as described above). Finally, we note that the actions produced by the non-interpolated St\"ackel fudge are, for all practical purposes, unbiased -- their average values along the orbit agree very well with calculations using a more rigorous generating function approach \citep{SandersBinney2014}.

%%%%%%%%%%%%%%%%%
\begin{lstlisting}[
caption={Action computation. \protect\\
We numerically integrate an orbit in an axisymmetric Miyamoto--Nagai potential for 10 dynamical times, and compute the actions at each point along the orbit, using the St\"ackel fudge or its interpolated version. They vary by $\lesssim 1\%$ even for this rather eccentric orbit ($J_r \simeq J_z \simeq J_\phi$). This script produces plots similar to Figure~\ref{fig:staeckel_orbit}.
}, label=code:actions, float=t]
import agama, matplotlib.pyplot as plt

# create an axisymmetric disc potential
pot = agama.Potential(type='MiyamotoNagai',
    scaleRadius=1, scaleHeight=0.2)

# integrate an orbit in this potential
initcond = [2, 0, 1.4, 0, 0.11, 0]  # x,y,z,vx,vy,vz
t, orb = agama.orbit(potential=pot, ic=initcond,
    time=10*pot.Tcirc(initcond), trajsize=1000)

# plot the orbit in the meridional plane (R,z)
plt.plot( (orb[:,0]**2 + orb[:,1]**2)**0.5, orb[:,2])
plt.show()

# construct action finders with or without interpolation
af_fudge  = agama.ActionFinder(pot, interp=False)
af_interp = agama.ActionFinder(pot, interp=True)

# compute the actions for each point along the orbit
act_f = af_fudge(orb)
act_i = af_interp(orb)

# plot the actions as a function of time
plt.plot(t, act_f[:,0], label='Jr, fudge')
plt.plot(t, act_f[:,1], label='Jz, fudge')
plt.plot(t, act_i[:,0], label='Jr, interp')
plt.plot(t, act_i[:,1], label='Jz, interp')
plt.plot(t, act_i[:,2], label='Jphi')
plt.legend()
plt.show()
\end{lstlisting}
%%%%%%%%%%%%%%%%%

To summarize, the axisymmetric St\"ackel fudge in \textsc{Agama} is both more efficient and accurate than other existing implementations, thanks to the new method for estimating the focal distance and various algorithmic improvements, and its interpolated version offers further significant speedup at the expense of an often tolerable decrease in accuracy. Listing~\ref{code:actions} illustrates the use of action finders.

%%%%%%%%%%%%%%%%%%%%%%%%%%%%%%%%
\section{Distribution functions}  \label{sec:df}

A stellar system consisting of a large enough number of stars generally can be described by a one-particle DF
$f(\boldsymbol{x},\boldsymbol{v},t)$, and according to Jeans' theorem, in a steady state it may only depend on the integrals of motion -- in our approach, actions. We use the convention for normalization of the DF such that its integral over some region of phase space equals the \textit{mass} (not \textit{number}) of stars in that region. Since the integration over angles is trivial, this mass is given by
\begin{align}  \label{eq:DFmass}
M=(2\pi)^3\iiint f(\boldsymbol{J})\,\mathrm{d}^3\!J,
\end{align}
without extra weight factors (such as the density of states $g(E)$, Equation~4.55 in \citealt{BinneyTremaine}) and independently of the potential.
In the core of the library we only deal with simple DFs of the form $f(\boldsymbol{J})$. More generally, the population of stars of various masses, ages and metallicities can be described by an extended DF that also depends on these extra arguments \citep{SandersBinney2015b}; such user-defined extended DFs can be used in the Python interface \citep[e.g.,][]{Das2016}. Alternatively, one may simply use a superposition of single-component DFs for each population \citep[e.g.,][]{Bovy2012,Trick2016}.

\textsc{Agama} provides several types of DFs for spheroidal and disky components, including generalizations of previously used models and some new ones. There are routines for computing DF moments (density, velocity dispersion, marginalized 1d velocity distributions, etc.), sampling a DF by particles, and determining the best-fit DF from discrete samples; some of them are illustrated in Listings \ref{code:df}, \ref{code:fit}.

%%%%%%%%%%%
\subsection{Double-power-law DF for spheroidal systems}  \label{sec:DFsphDPL}

A suitable choice for spheroidal components (bulge, halo) is a family of double-power-law DFs -- a generalization of the one introduced by \citet{Posti2015}:
\begin{align}  \label{eq:DFhalo}
f(\boldsymbol{J}) &= \frac{M}{(2\pi\, J_0)^3} 
\left[1 + \left(\frac{J_0}{h(\boldsymbol{J})}\right)^\eta \right]^{\Gamma/\eta}
\left[1 + \left(\frac{g(\boldsymbol{J})}{J_0}\right)^\eta \right]^{\frac{\Gamma-\mathrm{B}}{\eta}} \nonumber \\
&\times \exp\bigg[-\left(\frac{g(\boldsymbol{J})}{J_\mathrm{max}}\right)^\zeta\bigg]
\bigg(1 + \varkappa \tanh\frac{J_\phi}{J_{\phi,0}}\bigg).
\end{align}
Here
\begin{align*}
g(\boldsymbol{J}) &\equiv g_r J_r + g_z J_z\, + g_\phi\, |J_\phi|, \nonumber\\
h(\boldsymbol{J}) &\equiv h_r J_r + h_z J_z   + h_\phi   |J_\phi|  \nonumber
\end{align*}
are linear combinations of actions that control the flattening and anisotropy of the model in the outer region (above the break action $J_0$) and the inner region (below $J_0$), respectively. Only two of the three coefficients in the linear combination are independent, so we set $g_\phi=3-g_r-g_z$, $h_\phi=3-h_r-h_z$.
The power-law indices $\mathrm{B}$ and $\Gamma$ control the outer and inner slopes  of the density profile. The parameter $\eta$ determines the steepness of the transition between the two regimes. An optional exponential cutoff at $J \gtrsim J_\mathrm{max}$ is controlled by the sharpness parameter $\zeta$. The last term allows for rotation by introducing the odd-$J_\phi$ part of DF, proportional to the even part with a coefficient $\varkappa$ \citep{Binney2014}; $J_{\phi,0}$ then controls the extent of the central region where the rotation is suppressed. 

This DF roughly corresponds to a (possibly flattened) double-power-law density profile.
The asymptotic behaviour in a power-law regime at small or large radii can be derived analytically, 
yielding the values of power-law indices and mixing coefficients corresponding to the given density profile. 
\citealt{Posti2015} constructed DFs approximately corresponding to spherical double-power-law density profiles, by fixing these parameters to their asymptotic values, setting $\eta=1$ and adjusting the only free parameter $J_0$ to achieve the best agreement between the density profile generated by the DF (Section~\ref{sec:DFapplications}) and the required analytic density model. We argue that it's preferrable to freely adjust all DF parameters in order to attain a better match in the intermediate range of radii at the expense of a moderate deterioration in the asymptotic regime (e.g., reducing the approximation error in density and velocity dispersion from 10\% down to $\lesssim 1\%$ for a Hernquist model).
\citet{WilliamsEvans2015a} suggest a somewhat different functional form for the transition regime that can be better tuned to produce desirable anisotropy profiles. In the end, if the goal is to fit the DF parameters using observations (Section~\ref{sec:DFfitting}), it does not matter whether the DF produces a density profile that closely follows any particular functional form, so long as it adequately fits the observed kinematics and surface density.

%%%%%%%%%%%
\subsection{Quasi-isotropic DF for spheroidal systems}  \label{sec:DFsphIso}

An alternative way of constructing suitable DFs for spheroidal galaxy components is based on the following idea. For an arbitrary spherically-symmetric density profile $\rho(r)$ in a spherically-symmetric potential $\Phi(r)$ (not necessarily related to $\rho$ via Poisson equation), we numerically construct an isotropic DF of the form $f(E)$, using the Eddington inversion formula; this might be generalized to anisotropic models with $f(E,L)$ \citep[e.g.,][]{Cuddeford1991}. Even if there is no closed-form analytical expression for such a DF, it can always be approximated by a smooth and fast-to-compute function such as a cubic spline in suitably scaled variables.

In this spherical potential, we also know the expression for energy $E(J_r,L)$ as a function of actions $J_r$ and $L\equiv J_z+|J_\phi|$ (again, it may be approximated to any desired accuracy by a two-dimensional spline). Therefore, the DF may be viewed as a composition of two functions $f\big(E(J_r, L)\big)$, and the intermediate mapping $E(J_r, L)$ can be made part of its definition. In this interpretation, a `quasi-isotropic DF' $f(\boldsymbol{J})$ is a valid action-based DF.
Of course, it still has isotropic velocity distribution and produces the original density profile $\rho(r)$ in the same original potential $\Phi(r)$. But we may equally well use this DF in any other potential $\Phi'(\boldsymbol{r})$, not necessarily spherical. In this case, the DF may produce a non-spherical density profile and anisotropic kinematics (hence the `quasi-' name prefix).

The mapping between $\{\boldsymbol{x},\boldsymbol{v}\}$ and $\boldsymbol{J}$ depends on the potential, but the numerical value of $f(\boldsymbol{J})$ does not. Note that differs from computing the energy $E'\equiv \Phi'(\boldsymbol{r})+\frac{1}{2}|\boldsymbol{v}|^2$ and using the original function of energy $f(E')$ -- this would not be an action-based DF and hence would not enjoy its useful properties.
For instance, the density generated by a quasi-isotropic DF in the potential $\Phi'$ would be identical to the density obtained by evolving the population of stars drawn from the initial DF in a slowly varying potential that adiabatically changes from $\Phi$ to $\Phi'$ -- and this is achieved in one go, without the need to actually follow this evolution! (of course, this holds for any action-based DF, not only a quasi-isotropic one). This property is handy for the construction of multi-component self-consistent models, described in the \hyperref[par:advantages]{next section}.
The motivation is that if the final potential is not too different from the initial one (for instance, the initial potential is a spherically-symmetric version of the final one), then the resulting density profile would be similar to the initial one, have the same total mass, but somewhat flattened shape and moderately anisotropic kinematics.

A disadvantage of this type of DF is that it does not have a closed analytic form. One may wish to approximate such a numerically constructed DF by equation (\ref{eq:DFhalo}).
Among the example programs distributed with the library, we provide a tool for constructing such approximations for arbitrary spherical density and potential profiles (not necessarily related via Poisson equation).
\citet{Jeffreson2017} used a similar approach, but with a double-power-law DF of the form proposed by \citet{WilliamsEvans2015a}.

Thus we have two different approaches for defining a DF of a spheroidal component: one mandates that this DF produces a particular density profile in a particular potential (the quasi-isotropic DF), and the other dispenses with the mention of potential altogether, defining the functional form of the DF purely in terms of actions.
As we shall see, a similar distinction can be made for disky DFs.

%%%%%%%%%%%
\subsection{Quasi-isothermal DF for discs}  \label{sec:DFdiscIso}

Distribution functions for disky components are usually intended to produce density profiles $\rho(R,z) \simeq \Sigma(R)\,H(z)$, with the surface density $\Sigma(R)$ typically declining exponentially with radius, $\Sigma(R) = \frac{M_\mathrm{disc}}{2\pi\,R_\mathrm{disc}^2}\,\exp(-R/R_\mathrm{disc})$, and the vertical profile being isothermal, $H(z)\propto \mathrm{sech}^2\big(\frac{z}{2h}\big)$. In a kinematically cold disc, the velocity distribution at a radius $R$ is approximately Maxwellian with dispersions $\sigma_R(R),\sigma_z(R)$ that are small compared to the mean streaming velocity $\overline{v_\phi}$ (itself close to the circular velocity $v_\circ(R) \equiv \sqrt{R\,\partial\Phi/\partial R}$). Expressed in terms of integrals of motion, such a DF might look like
\begin{align}  \label{eq:DFshu}
f(E,L_z) = f_\circ(L_z)\,\exp\big(-E_R/\tilde\sigma_R^2-E_z/\tilde\sigma_z^2\big),
\end{align}
where $E_z \equiv \Phi(R,z)-\Phi(R,0)+v_z^2/2$ is the energy of vertical motion (an approximate integral),
$E_R \equiv E-E_\circ-E_z$ is the energy in planar motion, $E_\circ$ is the energy of a circular orbit with angular momentum $L_z$, and $f_\circ(L_z)$ is related to the radial disc profile. The functions controlling the velocity dispersions are tilded, to distinguish them from the true dispersions that are obtained by taking moments of the DF. This form is used by many authors, e.g., \citet{KuijkenDubinski1995}. To bring this DF into the action-based form, we replace $E_R$ and $E_z$ respectively with $\kappa J_r$ and $\nu J_z$, where
\begin{align*}
\kappa\equiv \sqrt{\frac{\partial^2\Phi}{\partial R^2} + \frac{3}{R}\,\frac{\partial\Phi}{\partial R}}, \quad \nu\equiv \sqrt{\frac{\partial^2\Phi}{\partial z^2}}
\end{align*}
are the radial and vertical epicyclic frequencies. The result is the quasi-isothermal DF of \citet{BinneyMcMillan2011}:
\begin{align}  \label{eq:DFbinney}
f(\boldsymbol{J}) = f_\circ(J_\phi) \times
\frac{\kappa}{\tilde\sigma_R^2}\exp\left(-\frac{\kappa J_r}{\tilde\sigma_R^2}\right) \times
\frac{\nu   }{\tilde\sigma_z^2}\exp\left(-\frac{\nu    J_z}{\tilde\sigma_z^2}\right). \hspace{-1cm}
\end{align}
Here the frequencies and velocity dispersions must be functions of actions; again the simplest way to ensure this is to let them depend on the radius $R_\circ$ of a circular orbit with angular momentum $J_\phi$, defined as the solution of the equation $R^3\,\partial\Phi/\partial R = J_\phi^2$.
The approximate relation between $f_\circ(J_\phi)$ and the surface brightness profile $\Sigma(R)$ may be derived by integrating the above expression over angles and actions and noting that the integral over $J_\phi$ may be converted into the integral over $R_\circ(J_\phi)$: 
\begin{align*}
M &= \int f(\boldsymbol{J})\;\mathrm{d}^3J\,\mathrm{d}^3\theta =
(2\pi)^3 \int f_\circ(J_\phi)\, \mathrm{d}J_\phi \\
&= (2\pi)^3 \int f_\circ(J_\phi)\, \frac{\mathrm{d}J_\phi}{\mathrm{d}R_\circ}\, \mathrm{d}R_\circ \\
&= (2\pi)^3 \int f_\circ(J_\phi)\, \frac{R_\circ\,\kappa^2}{2\Omega}\, \mathrm{d}R_\circ
\end{align*}
where $\Omega\equiv v_\circ/R$ is the azimuthal epicyclic frequency.
On the other hand, the same mass is given by
\begin{align*}
M &= \int \Sigma(R_\circ)\,2\pi\,R_\circ\,\mathrm{d}R_\circ.
\end{align*}
Comparing the two expressions, we infer a suitable form of $f_\circ(J_\phi)$:
\begin{align}
f_\circ(J_\phi) &= \frac{\tilde\Sigma(R_\circ)\,\Omega(R_\circ)}{2\pi^2\,\kappa^2(R_\circ)} , \label{eq:DFdisk_Jphi}
\end{align}
with the function $\tilde\Sigma(R_\circ)$ matching the surface brightness profile $\Sigma(R)$.

This DF is not without deficiencies. First, in the case when the potential is cuspy (e.g., a Hernquist bulge), the epicyclic frequencies rise arbitrarily high at small radii, so this DF is not well-behaved at small $J_\phi$. More generally, for a realistically warm disc there is a significant fraction of stars with $\mathrm{max}(J_r,J_z) \gtrsim J_\phi$ (especially at small radii), for which the dimensional quantities entering the DF (density, frequencies, velocity dispersions) are evaluated at a radius $R_\circ(J_\phi)$ that is much smaller than the average radius of the orbit. As a result, the density $\Sigma$ and velocity dispersions $\sigma_R,\sigma_z$ generated by such a DF are not particularly close to their desired forms $\tilde\Sigma,\tilde\sigma_R,\tilde\sigma_z$ -- usually they have a pronounced central depression at radii smaller than the disc scale length.

\citet{Dehnen1999} proposed a modification of this approach: the radius used to compute the surface density, velocity dispersions and epicyclic frequencies, is taken to be the radius of a circular orbit with the given \textit{energy}, not \textit{angular momentum}: he argues that it corresponds more closely to the actual mean radius of an orbit even if the latter is far from circular.
In terms of actions, this may be mimicked by replacing $R_\circ(J_\phi)$ with $R_\circ(\tilde J)$, where
\begin{align}  \label{eq:DFdisk_Jsum}
\tilde J \equiv |J_\phi| + k_r\,J_r + k_z\,J_z
\end{align}
is a linear combination of actions that is approximately constant across the energy surface (see Figure~1 in \citealt{Binney2014} for an illustration). For orbits close to the disc plane, this could be achieved by setting $k_r = \kappa/\Omega$, $k_z = \nu/\Omega$, but more generally these coefficients may have arbitrary values. Larger $k_z$ produce DF that declines faster at large $J_z$, suppressing the high-$z$ tail of the vertical density profile or the high-$v_z$ tail of the velocity distribution, without noticeably changing its width, and $k_r$ has a similar effect on the radial velocity distribution. Values of order unity produce a suitable DF which generates a density profile that more closely follows the desired form $\tilde\Sigma(R)$ than an un-modified DF.

It remains to choose the form of functions $\tilde\sigma_R(R_\circ)$, $\tilde\sigma_z(R_\circ)$.
The vertical velocity dispersion is often taken to be exponentially declining with radius:
\begin{subequations}  \label{eq:DFdisk_sigmaz}
\begin{align}  \label{eq:DFdisk_sigmaz_exp}
\tilde\sigma_z = \sigma_{z,0}\, \exp(-R_\circ/R_{\sigma_z}),
\end{align}
because this would produce a constant scale height in an isolated thin disc. In the presence of other components, the self-gravity of the disc is not necessarily dominant at small and large radii, so a better approximation to a constant-scaleheight disc is obtained by setting 
\begin{align}  \label{eq:DFdisk_sigmaz_iso}
\tilde\sigma_z(R_\circ) = \sqrt{2}\,h\,\nu(R_\circ).
\end{align}
\end{subequations}
Similarly, the radial velocity dispersion is also assumed to decline exponentially with radius (for no other reason than similarity to $\tilde\sigma_z$):
\begin{align}  \label{eq:DFdisk_sigmaR}
\tilde\sigma_R(R_\circ) = \sigma_{R,0}\, \exp(-R_\circ / R_{\sigma_R}), 
\end{align}
with adjustable maximum value $\sigma_{R,0}$ and scale length $R_{\sigma_R} \simeq 2R_\mathrm{disc}$.
We retain this functional form, but place a lower limit on how small $\tilde\sigma_R$ could get, to avoid an unphysical situation that the DF in the disc plane could become arbitrarily high at large radii.

Finally, the disc DF should normally be asymmetric in $J_\phi$, producing net rotation. \citet{BinneyMcMillan2011} achieve this by multiplying the DF by a factor $1 + \mathrm{tanh}(J_\phi/J_{\phi,0})$, with $J_{\phi,0}$ chosen such that in the central part of the disc there is approximate symmetry between prograde and retrograde orbits, and further out only the prograde ones survive.
Following \citet{Dehnen1999}, we instead introduce an alternative method for extending the DF into the negative-$J_\phi$ half-space, namely, multiply $f$ by 
\begin{align}  \label{eq:DFdisk_negJphi}
f_{\pm} \equiv \left\{ \begin{array}{ll}
1 &\mbox{if }J_\phi\ge 0 , \\
\exp\big( 2\Omega J_\phi / \sigma_R^2 \big) &\mbox{if }J_\phi<0 ,
\end{array} \right.
\end{align}
where $\Omega$ and $\sigma_R$ are evaluated at the radius $R_\circ(\tilde J)$.
To summarize, the improved variant of quasi-isothermal DF is given by
\begin{align}  \label{eq:DFdisk_quasiiso}
f(\boldsymbol{J}) &= \frac{\Sigma_0\,\Omega(R_\circ)}{2\pi^2\,\kappa(R_\circ)^2}
\exp\bigg(-\frac{R_\circ}{R_\mathrm{disc}}\bigg) \times f_\pm \\
&\times \frac{\kappa(R_\circ)}{\tilde\sigma_R^2(R_\circ)}
\exp\bigg(-\frac{\kappa(R_\circ)\,J_r}{\tilde\sigma_R^2(R_\circ)}\bigg) \nonumber\\
&\times \frac{\nu(R_\circ)}{\tilde\sigma_z^2(R_\circ)}
\exp\bigg(-\frac{\nu(R_\circ)\,J_z}{\tilde\sigma_z^2(R_\circ)}\bigg), \nonumber
\end{align}
where $R_\circ$ is a function of $\tilde J$ (\ref{eq:DFdisk_Jsum}), velocity dispersions are given by (\ref{eq:DFdisk_sigmaz},\ref{eq:DFdisk_sigmaR}), and $f_\pm$ is given by (\ref{eq:DFdisk_negJphi}).

If the disc is not too cold, the density and velocity dispersions generated by such a DF may be different from the required values. \citet{KuijkenDubinski1995}, \citet{Dehnen1999} or \citet{Sharma2013} further adjust the functions $\tilde\Sigma(R),\tilde\sigma_{R,z}(R)$ to match the resulting profiles more closely to the input ones. We do not employ such adjustments, relying instead on the recomputation of the potential to achieve self-consistency, as described in Section~\ref{sec:scm_present}.
Our modifications -- the use of a linear combination of actions (\ref{eq:DFdisk_Jsum}) and a different expression for $\tilde\sigma_z$ (\ref{eq:DFdisk_sigmaz_iso}) -- generally produce a better match to the density profile of a radially-exponential, vertically-isothermal disc, than the original quasi-isothermal DF of \citet{BinneyMcMillan2011} (which could still be recovered by using Equation~\ref{eq:DFdisk_sigmaz_exp} and setting $k_r=k_z=0$).

%%%%%%%%%%%%%%%
\begin{figure*}
\includegraphics{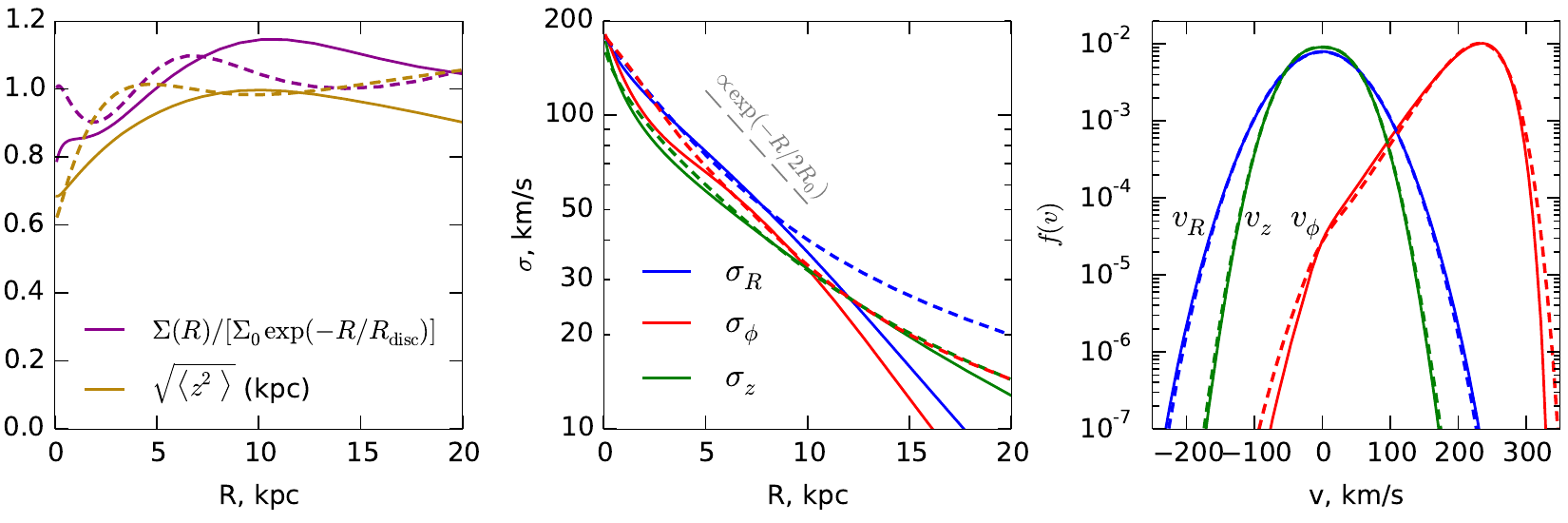}
\caption{Illustration of disc distribution functions. We take the Milky Way potential from \citet{Piffl2014} and consider two types of DFs: quasi-isothermal (solid lines) and exponential (dashed), designed to produce a realistic thick disc population (scale radius $\sim 3$~kpc, scale height $\sim 1$~kpc, radial and vertical velocity dispersions at $R=8$~kpc equal to $\sigma_R=50$~km/s, $\sigma_z=40$~km/s, consistent with the values given in \citealt{BlandHawthornGerhard2016}). The DF parameters are: $R_\mathrm{disc}=3$~kpc, $\sigma_{R,0}=185$~km/s, $R_{\sigma_R}=2R_\mathrm{disc}$, $h=0.38$~kpc for the quasi-isothermal DF (\ref{eq:DFdisk_sigmaz_iso}--\ref{eq:DFdisk_quasiiso}), or $\{J_{r,0}, J_{z,0}, J_{\phi,0}, J_{d,0}, J_{v,0}\} = \{340, 260, 720, 500, 720\}$~kpc$\times$km/s for the exponential DF (\ref{eq:DFexp}); the mixing coefficients in (\ref{eq:DFdisk_Jsum}) are $k_r=1, k_z=0.25$ in both cases. \protect\\
Left panel shows the radial dependence of the surface density, normalized to the exponential profile $\Sigma_0\exp(-R/R_\mathrm{disc})$, and the r.m.s. value of $z$, expected to be nearly constant. The actual profiles do not exactly follow the expectations, but the deviations are within $10-20\%$, except at the very centre, where the disc is no longer a dominant component. The vertical density profile (not shown) very closely follows the sech${}^2$ law. \protect\\
Central panel plots the radial dependences of three components of the velocity dispersion tensor in the equatorial plane. The vertical velocity dispersion $\sigma_z$ is computed from the condition that the scale height is (nearly) constant, and follows very similar profiles for both DF models, but is not exponentially declining with radius, as is commonly assumed (if it were, the scale height would not stay constant). The radial velocity dispersion $\sigma_R$ is a nearly exponential function of radius in the quasi-isothermal DF, and consequently falls below the vertical one at large radii, which is not very realistic. In the exponential DF it behaves similarly to $\sigma_z$. \protect\\
Right panel shows the 1d velocity distribution functions $\mathfrak{f}(v)$ for each velocity component, at a point $R=8$~kpc, $z=0$. They are very similar in both models, nearly gaussian for $v_R, v_z$ and strongly asymmetric for $v_\phi$, peaking around the local circular velocity (240 km/s), shaprly declining towards larger $v_\phi$ and having a long tail extending to negative $v_\phi$, described by Equation~\ref{eq:DFdisk_negJphi} (although in a realistic galaxy model, the halo stellar population dominates at these velocities).
}  \label{fig:df_moments}
\end{figure*}
%%%%%%%%%%%%%

As in the case of quasi-isotropic DFs, the intermediate mappings $J_\phi \to R_\circ \to \kappa,\nu,\Omega,\Sigma$, etc., become part of the definition of the DF, constructed in a particular potential, but not necessarily the one that the DF is used in later. Of course, if the potential is very different from the one used to initialize these mappings, then the density generated by such a DF will also differ significantly from the exponential/isothermal form.

%%%%%%%%%%%
\subsection{Exponential DF for discs}  \label{sec:DFdiscExp}

The fact that the DF contains functions defined by a potential is conceptually distressing, and we now introduce another family of disc DFs that is specified entirely in terms of actions, without any auxiliary functions. It is designed to produce density profiles that are close to exponential/isothermal, in a potential with an approximately flat rotation curve ($v_\circ\approx \mathrm{const}$).
In this case, $R_\circ(J_\phi) = |J_\phi| / v_\circ$, $\Omega = v_\circ^2 / |J_\phi|$, $\kappa = \sqrt{2}\,\Omega$, and $\nu$ is also approximately proportional to $\Omega$, although with a coefficient that slowly varies with radius. Velocity dispersions are related to the extent of radial ($\Delta R$) and vertical ($h$) excursions of stars: $\sigma_R\sim \kappa\, \Delta R$, $\sigma_z\sim \nu\, h$, and characteristic values of actions are, correspondingly, $\langle J_r\rangle \sim \sigma_R\Delta R \sim \kappa\Delta R^2$, $\langle J_z\rangle \sim \sigma_z h \sim \nu z^2$. The dependence of DF on the vertical action should be exponential with the characteristic scale $\langle J_z\rangle$:
\begin{align*}
f \propto \exp\bigg( -\frac{J_z}{\langle J_z \rangle} \bigg) \sim
\exp\bigg( -\frac{J_z |J_\phi|}{(v_\circ h)^2} \bigg) =
\exp\bigg( -\frac{J_z |J_\phi|}{J_{z,0}^2} \bigg),
\end{align*}
where we introduced a new dimensional parameter $J_{z,0} \equiv v_\circ h$.
Similar considerations lead to the same exponential dependence of DF on the radial action, with its own characteristic scale $J_{r,0}$. Finally, the analog of (\ref{eq:DFdisk_Jphi}) for an exponential radial density profile would be $f_\circ(J_\phi) \propto |J_\phi| \exp( - |J_\phi| / J_{\phi,0})$.
In practice, these expressions need to be tweaked somewhat: first, we replace $|J_\phi|$ by a linear combination $\tilde J$ of all three actions (\ref{eq:DFdisk_Jsum}), for the same reasons as in the quasi-isothermal DF (it better corresponds to the average radius of the orbit even when the latter is far from circular). Second, recognizing that the potential cannot have a flat rotation curve all the way down to $R\to 0$, we mitigate undesirable deviations of the model from its prescribed properties at small radii by two extra parameters with the dimension of actions $J_{d,0}$ and $J_{v,0}$.
In the end, the new `exponential' DF model is defined as follows:
\begin{align}  \label{eq:DFexp}
f(\boldsymbol{J}) &= \frac{M}{(2\pi)^3}\,
\frac{J_d}{J_{\phi,0}^2}\, \exp\bigg(-\frac{J_d}{J_{\phi,0}}\bigg) \\
&\times \frac{J_v}{J_{r,0}^2}\, \exp\bigg(-\frac{J_v\,J_r}{J_{r,0}^2}\bigg)
 \times \frac{J_v}{J_{z,0}^2}\, \exp\bigg(-\frac{J_v\,J_z}{J_{z,0}^2}\bigg) \nonumber\\
&\times\left\{\begin{array}{ll} 
1 &\mbox{if } J_\phi \ge 0 \;, \\
\exp\bigg(\displaystyle\frac{J_v\,J_\phi}{J_{r,0}^2} \bigg) &\mbox{if } J_\phi < 0 \;,
\end{array}\right. \nonumber  \\
J_d &\equiv \sqrt{\tilde J^2 + J_{d,0}^2}, \quad
J_v  \equiv \sqrt{\tilde J^2 + J_{v,0}^2}. \nonumber
\end{align}

The advantage of this DF is a relatively simple functional form, which produces disky density profiles in realistic potentials (not necessarily with flat rotation curves), and a small number of free parameters: $M$ sets the overall normalization (the total mass is of order $M$), $J_{\phi,0}$ determines the scale radius, $J_{z,0}$ -- scale height and vertical velocity dispersion, $J_{r,0}$ -- radial velocity dispersion, $J_{d,0} \simeq J_{r,0}$ affects the density profile at small radii, and $J_{v,0} \simeq J_{\phi,0}$ influences the velocity distribution in the centre.

Thus we have two alternative models for the disc DF -- the quasi-isothermal one is specified in terms of parameters with dimensions of length and velocity, and needs a potential model to construct auxiliary functions, whereas the exponential DF is specified entirely in terms of scale actions and does not reference any potential. This parallels the distinction between quasi-isotropic and double-power-law DFs for spheroidal components discussed in the previous section. Depending on the application, one or the other approach may be more convenient. Figure~\ref{fig:df_moments} illustrates that both families of disc DFs produce similar and realistic density profiles and velocity distributions in a typical Milky Way potential.

%%%%%%%%%%%
\subsection{Using the DF}  \label{sec:DFapplications}

Given the DF and the potential (which determines the mapping $\boldsymbol{J}[\boldsymbol{x},\boldsymbol{v}]$), we may compute various moments by integrating the DF over velocity:
\begin{align}
\rho(\boldsymbol{x}) &\equiv \iiint \mathrm{d}^3v\; f\big(\boldsymbol{J}[\boldsymbol{x},\boldsymbol{v}]\big)
&&\mbox{density}, \label{eq:DensityMoment} \\
\overline{\boldsymbol{v}} &\equiv \frac{1}{\rho} \iiint \mathrm{d}^3v\; \boldsymbol{v}\; f\big(\boldsymbol{J}\big)
&& \mbox{mean velocity}, \nonumber\\
\overline{v^2_{ij}} &\equiv \frac{1}{\rho} \iiint \mathrm{d}^3v\; v_i v_j\; f\big(\boldsymbol{J}\big) 
&& \mbox{second moment of velocity}, \nonumber\\
\sigma^2_{ij} &\equiv \overline{v^2_{ij}} - \overline{v_{i}}\,\overline{v_{j}}
&& \mbox{velocity dispersion tensor}.\nonumber
\end{align}
A more complete characterization of kinematic structure is given by the 1d distribution of any velocity component $v_1$, marginalized over the other two components $v_2,v_3$:
\begin{align}  \label{eq:veldistrib}
\mathfrak{f}(\boldsymbol{x};\,v_1) \equiv \frac{1}{\rho} \iint \mathrm{d}v_2\,\mathrm{d}v_3\;f\big(\boldsymbol{J}\big) .
\end{align}
We compute these integrals by an adaptive multidimensional routine \texttt{cubature}. %Velocity distributions are represented as B-splines of degree from 0 to 3: degree 0 results in a simple histogram (piecewise-constant $\mathfrak{f}$), while degree 3 is a familiar cubic spline. All coefficients of this B-spline representation are computed simultaneously in the course of a single integration of a multi-valued function.
Listing~\ref{code:df} illustrates the computation of DF moments and velocity distributions.

%%%%%%%%%%%%%%%%%
\begin{lstlisting}[
caption={Computation of DF moments. \protect\\
We take a realistic Milky Way potential \citep{Piffl2014} and construct a quasi-isothermal DF with parameters corresponding to the thick disc. Combination of a DF and a potential, together with an implicitly created action finder, represents a model for a single stellar population in the galaxy; this object provides methods for computing various DF moments and sampling from the DF.
We plot the radial profiles of density and velocity dispersions (\ref{eq:DensityMoment}) in the equatorial plane, and the 1d velocity distributions (\ref{eq:veldistrib}) at a particular point; this roughly reproduces Figure~\ref{fig:df_moments}.
}, label=code:df, float=t]
import agama, numpy, matplotlib.pyplot as plt

# assign the units (1 Msun, 1 Kpc, 1 km/s)
agama.setUnits(mass=1, length=1, velocity=1)

# load one of the standard Milky Way potential models
pot = agama.Potential('Piffl14.ini')

# initialize the distribution function of a thick disc
df  = agama.DistributionFunction(type='QuasiIsothermal',
    Sigma0=1e3, Rdisk=3, hdisk=0.5,
    sigmaR0=200, RsigmaR=6, potential=pot)

# bind together the potential and the DF
mod = agama.GalaxyModel(pot, df)

# choose the range of radii to plot DF moments
r   = numpy.linspace(0.0, 15.0, 30)
xyz = numpy.column_stack((r, r*0, r*0))

# compute the density and velocity dispersions at these radii
dens,vel2 = mod.moments(xyz)
plt.plot(r, dens, label='rho(R, z=0)')
plt.plot(r, vel2[:,0]**0.5, label='sigma_R')
plt.plot(r, vel2[:,1]**0.5, label='sigma_z')
plt.yscale('log')
plt.legend()
plt.show()

# compute the 1d velocity distributions at a single point
vdfr, vdfz, vdfphi = mod.vdf([8, 0, 0])  # x,y,z
vgrid = numpy.linspace(-200, 320, 250)   # grid in velocity
plt.plot(vgrid, vdfr(vgrid), label='f(v_r)')
plt.plot(vgrid, vdfz(vgrid), label='f(v_r)')
plt.plot(vgrid, vdfphi(vgrid), label='f(v_phi)')
plt.yscale('log')
plt.legend(loc='lower right')
plt.show()
\end{lstlisting}
%%%%%%%%%%%%%%%%%

The DF can be sampled with particles in the 3d action space, or, more usefully in practice, in the 6d position/velocity space -- the latter could be used to create initial conditions for $N$-body simulations. We employ an adaptive multidimensional rejection sampling algorithm, which iteratively refines regions in the sampling domain where the function values are highest. It is completely general (does not use any prior information about the function), quite efficient (acceptance fraction is typically $\gtrsim 10-30\%$ even for a strongly localized function), and can be used in many other contexts. For instance, a mock catalogue of stars for a particular survey is produced by sampling from a DF multiplied by this survey's selection function.

%%%%%%%%%%%
\subsection{Fitting a DF}  \label{sec:DFfitting}

The reverse scenario is to use a sample of observed stars to infer the parameters of the galaxy model $M$ -- the  potential and the DF \citep[e.g.,][]{McMillanBinney2013,BovyRix2013}. Suppose that we know the survey's selection function $S(\boldsymbol{w})$, where $\boldsymbol{w}$ are the observationally determined quantities (e.g., sky coordinates, parallax, proper motion, line-of-sight velocity, magnitude, etc.); this function describes how likely we are to observe a star with these parameters if it actually exists in the galaxy. We know the mapping between $\boldsymbol{w}$ and the phase-space coordinates $\{\boldsymbol{x},\boldsymbol{v}\}$, and in a given potential, also know how they translate into actions $\boldsymbol{J}$. Since these quantities are not measured exactly, and some of them may not be accessible at all, we need to consider the multivariate error distribution $E(\boldsymbol{w} | \boldsymbol{w}')$, describing the probability of measuring $\boldsymbol{w}$ given the true values $\boldsymbol{w}'$. Then the likelihood of observing a star $s$ with coordinates $\boldsymbol{w}_s$, given the model $M$ and the survey $S$, is 
\begin{align}  \label{eq:Likelihood}
\mathcal{L}_s = \frac{ S(\boldsymbol{w}_s)\; \int \mathrm{d} w'\; E(\boldsymbol{w}_s | \boldsymbol{w}')\; f\big(\boldsymbol{J}[\boldsymbol{w}']\big)\; \mathcal{J}(\boldsymbol{w}') }
{ \int \mathrm{d} w'\; S(\boldsymbol{w}')\; f\big(\boldsymbol{J}[\boldsymbol{w}']\big)\;
\mathcal{J}(\boldsymbol{w}') } .
\end{align}
Here $f$ is the DF of stars in the model, $\mathcal{J} \equiv |\partial(\boldsymbol{x},\boldsymbol{v})/\partial\boldsymbol{w}|$ is the Jacobian of coordinate transformation, the integral in the numerator is the convolution with error distribution, and the integral in the denominator is the normalisation factor (the total number of stars that we expect to find in the survey for the given model).
In practice, the former integral is computed in a Monte Carlo way, by drawing $K$ sample points $\boldsymbol{w}_{s,k}$ from the (known) multivariate error distribution of observables for each star $s$, propagating them to action space, and replacing the integral by a sum
$\frac{1}{K} \sum_{k=1}^K f\big(\boldsymbol{J}[\boldsymbol{w}_{s,k}]\big)$.
It could well be that some of these points are unphysical (e.g., have negative parallax, or velocity exceeding the escape velocity); they will not contribute to the integral because $f=0$ for such points. As long as $f$ is non-zero for some points, the likelihood of a star remains positive.
The integral in the denominator is the same for all stars, and it should be computed accurately \citep[e.g.,][Section 2.6]{Trick2016}; we again use the adaptive \texttt{cubature} routine.
Finally, the log-likelihood of the model is given by summing the log-likelihoods of all stars in the observed sample. The model parameters can then be varied to maximize this quantity. In doing so, it is advisable to fix the Monte Carlo samples used to compute the error convolution for each star, minimizing the impact of Poisson noise \citep{McMillanBinney2013}.

%%%%%%%%%%%%%%%%%
\begin{lstlisting}[
caption={Fitting a DF model to data.\protect\\
We construct a mock catalogue of stars by sampling from a given DF in the given potential, and then try to recover the parameters of both the DF and the potential using the maximum-likelihood approach. In this simplified example, we vary only two parameters of the potential (mass and scale radius of a Plummer model) and one parameter of the double-power-law DF (\ref{eq:DFhalo}) (scale action $J_0$). We also assume that all 6 phase-space coordinates for all stars are known exactly, and ignore the selection function. For each trial choice of parameters, we compute the actions for all stars from the mock catalogue. The likelihood for star $s$ with coordinates $\boldsymbol{w}_s$ is given by (\ref{eq:Likelihood}), where the numerator is just 
$f(\boldsymbol{J}\lbrack\boldsymbol{w}_s\rbrack)$
and the denominator is the total mass associated with the DF (\texttt{norm}). The log-likelihood of the entire sample is the sum of log-likelihoods for each star; if any of them has positive energy in the trial potential, this potential has to be rejected. Finally, we use the minimization routine from \texttt{scipy} to find the minimum of negative log-likelihood. After $\sim100$ steps it converges to the true parameters used to generate the mock sample.
}, label=code:fit, float=t]
import agama, numpy
from scipy.optimize import minimize

# create a mock catalogue of stars (x,v)
truepot = agama.Potential(type='Plummer',
    scaleRadius=1, mass=1)
truedf = agama.DistributionFunction(type='DoublePowerLaw',
    norm=1, slopeIn=1, slopeOut=5, J0=1)
data,_ = agama.GalaxyModel(truepot, truedf).sample(10000)

# function to minimize is minus log-likelihood
def modelSearchFnc(params):
    # two parameters for the potential and one for the DF
    R, M, J0 = numpy.exp(params)
    pot = agama.Potential(type='Plummer',
        scaleRadius=R, mass=M)
    df = agama.DistributionFunction(type='DoublePowerLaw',
        norm=1, slopeIn=1, slopeOut=5, J0=J0)
    norm = df.totalMass()
    # compute actions for the mock samples in this potential
    ac = agama.actions(data, pot)
    # compute log-likelihood of these samples given the DF
    logL = numpy.sum( numpy.log( df(ac) / norm ) )
    # if any star has positive energy, this model is excluded
    if numpy.isnan(logL): logL = -numpy.inf
    print("scaleRadius=%f, mass=%f, J0=%f => logL=%f" %
        (R, M, J0, logL))
    return -logL

# start the search from arbitrary (wrong) initial values
initvals = numpy.array([2., 3., 4.])
minimize(modelSearchFnc, initvals, method='Nelder-Mead')
\end{lstlisting}
%%%%%%%%%%%%%%%%%

Depending on the context, we may keep either the potential or the DF fixed; for instance, if the goal is to infer the parameters of the potential, we should consider all possible combinations of the potential and the DF, and compute the posterior probability distributions of potential parameters, marginalized over the DF parameters. Listing~\ref{code:fit} illustrates the simplest scenario with full phase-space information and no errors. The true parameters of the potential and the DF are easily recovered using deterministic optimization routine; in practice this should be followed or replaced by a full exploration of parameter space using MCMC or similar aproaches. A more realistic scenario of inferring the potential from a catalogue of tracer stars with only sky position and noisy line-of-sight velocity (no distance or proper motion information), but still assuming a spherical geometry and neglecting the selection function, was applied to the Gaia Challenge test suite (Read et al., in prep.); the example code is provided with the library.

A particular case of inferring the DF from discrete samples occurs in the context of spherical isotropic DFs constructed from $N$-body snapshots. Here the DF is a function of one variable (energy, or rather its action-like counterpart -- phase volume), and the method of constructing a non-parametric spline representation of a one-dimensional probability distribution from discrete samples is based on penalized log-density estimate. It is used in the Monte Carlo code \textsc{Raga} to compute two-body relaxation coefficients.

%%%%%%%%%%%%%%%%%%%%%%%%%%%%%%%%
\section{Self-consistent models}  \label{sec:scm}

We now consider the task of constructing a galaxy model consisting of an arbitrary number of components (e.g., disc, bulge, halo), each specified by its DF $f_c(\boldsymbol{J})$, plus optionally some external potential (e.g., a central black hole). The problem is to find the solution of the coupled system of equations:
\begin{itemize}
\item actions $\boldsymbol{J}(\boldsymbol{x}, \boldsymbol{v}\,|\,\Phi)$ depend on the potential;
\item density of each component is the integral of its DF (\ref{eq:DensityMoment});
\item the potential is related to the total density via the Poisson equation (\ref{eq:Poisson}).
\end{itemize}

Because of this circular dependency, the solution is obtained iteratively, starting from a suitable initial guess for the potential and repeating the three steps until convergence. 
Our approach extends previous works, described in the next section, and improves them in several ways:
\begin{itemize}
\item we always recompute the density from the DF on each iteration, instead of relying on the assumption that the DF (e.g., a quasi-isothermal one) approximately generates the desired density profile;
\item consequently, we recalculate the total potential of all components without any simplifying approximations (e.g., of a separable exponential disc profile);
\item the use of actions facilitates the construction of models with arbitrary combination of components and accelerates the convergence of iterations.
\end{itemize}

%%%%%%%%%%%
\subsection{Previous work}  \label{sec:scm_previous}

The first application of this method dates to \citet{PrendergastTomer1970}, who constructed models of elliptical galaxies specified by $f(E,L_z)$, using spherical-harmonic potential representation; they obtained solution in a few dozen iterations, but not for every possible choice of DF (very flattened models could not be constructed with their family of DFs).
\citet{Rowley1988} used a similar approach, with a different ansatz for $f(E,L_z)$, and added a static disc potential (without a corresponding DF).

\citet{KuijkenDubinski1995} extended the method to the case of three-component disc galaxies with a bulge and a halo. They start from suitable ansatzes for the DFs of spheroidal components, expressed as analytic functions of $E$, $L_z$, and a static potential of the disc. Then they employ the same iterative procedure to recompute the total potential, using the modification of spherical-harmonic expansion for separable discs, described in Section~\ref{sec:potential}. Finally, the disc DF is constructed in the form given by (\ref{eq:DFshu}), using the energy of vertical motion $E_z$ as the approximate third integral. The functions $f_\circ(R_\circ)$, $\tilde\sigma_R(R_\circ)$, $\tilde\sigma_z(R_\circ)$ are initially taken to be exponential with radius, and are iteratively adjusted until the density matches the desired form (exponential in radius, isothermal vertically) at some pre-selected points; they argue that this produces the model sufficiently close to equilibrium so that the disc density does not need to be recomputed in the iteratively adjusted potential. Apart from this assumption, another difficulty is that the DFs for spheroidal components only produce their intended density profiles in isolation, but not in the composite model. This complicates the setup, because the relation between DF parameters and the resulting density profiles is rather non-intuitive.

\citet{Widrow2008} improved the method by using the Eddington inversion formula to numerically construct DFs of spheroidal components in the \textit{spherical approximation} of the total potential (retaining only the monopole term of the disc potential). Then they kept these expressions for $f(E)$, substituting $E$ with the energy computed in the actual potential at each iteration and rescaled back into the same range as used in the original DF. This approach resembles the quasi-isotropic DF introduced in Section~\ref{sec:DFsphIso}, but with important differences: the energy is not an adiabatic invariant, even after rescaling, and the total mass of such a DF depends on the potential even if the functional form $f(E)$ remains unchanged. In the final model the density of spheroidal components is constant on equipotential surfaces, i.e. is somewhat flattened compared to the initial profiles, but follows approximately the same radial dependence; the velocity distribution is necessarily isotropic.
This method, dubbed \textsc{GalactICs}, has been used to construct initial conditions for $N$-body simulations, as well as models of actual galaxies (Milky Way, M31) using a variety of observational constraints, and more recently also incorporating integral-field kinematic data for more distant galaxies \citep{Taranu2017}.

\citet{DebattistaSellwood2000} constructed initial conditions for their disc+halo $N$-body simulations using a similar iterative technique, recomputing the density profile of the halo from its DF (also expressed in terms of $E$ and $L_z$) and then readjusting the potential. For the latter step, they employed the same cartesian-grid Poisson solver as used in the actual $N$-body simulation, rendering unnecessary the modification of spherical-harmonic expansion used by \citet{KuijkenDubinski1995}. However, they also did not recompute the disk density profile, drawing particles from an approximately isothermal DF at the end of iterative procedure.

\citet{McMillanDehnen2007} used a somewhat different approach to construct equilibrium multicomponent models, implemented in a program \textsc{mkgalaxy} included in the \textsc{Nemo} framework. They also start from a spherical model with DFs of bulge and halo computed from 
\defcitealias{Cuddeford1991}{Cuddeford's (1991)}\citetalias{Cuddeford1991} anisotropic generalization of Eddington formula in the spherical approximation of the total potential (including the monopole component of the disc). Then the two spheroidal components are converted to $N$-body representations, and the potential of the disc component (which is still an external contribution to the $N$-body system of bulge and halo) is slowly deformed into the final shape, while the density profiles of the other two components adiabatically evolve while retaining overall self-consistency throughout the $N$-body simulation. Finally, disc particles are drawn from the DF of \citet{Dehnen1999}, in which all relevant quantities (circular radius corresponding to the given energy, epicyclic frequencies as functions of radius, etc.) are computed in the final potential, and the auxiliary functions $f_\circ(R)$, $\tilde\sigma_R(R)$ are also adjusted to produce the required density and velocity dispersion profiles (similarly to \citet{KuijkenDubinski1995}). The vertical velocity dispersion is taken from an isolated exponential disc, resulting in an out-of equilibrium vertical structure, which should probably be regarded as an oversight rather than an intrinsic deficiency of the method. This approach circumvents the need for iterations, replacing it with an actual adiabatic transformation of the distribution of particles (which is, in fact, more computationally intensive); however, it also forfeits the description of the bulge and halo in terms of a DF.

In these approaches the DF is represented as a function of energy $E$, angular momentum $L$ or $L_z$, and, for the disc component, an approximate third integral (the energy of vertical motion $E_z$). However, these variables are not ideal for iterative self-consistent modelling, whereas actions have several important advantages:
\begin{itemize}  \phantomsection\label{par:advantages}
\item A DF specified in terms of energy is inconvenient because the possible range of energy depends on the overall potential. Moreover, even the total mass of each component is hard to specify in advance. In the method of \citet{Widrow2008}, the DF is rescaled at each iteration, which probably explains slow convergence (several tens of iterations are needed, with the order of spherical-harmonic expansion increased gradually).
By contrast, a DF given in terms of actions always has the same mass (\ref{eq:DFmass}) and functional form, independently of the potential.
\item A superposition of several such components is straightforward, and keeps unchanged the explicit form of the DF in terms of $\boldsymbol{J}$. Physically this corresponds to adiabatic modification of the density profile of each component upon the addition of another component.
\item The vertical energy is a good proxy for the integral of motion only for cold orbits near the disc plane (where the epicyclic approximation is valid). The third integral in the St\"ackel approximation \citep[e.g.][]{Bienayme2015} is conserved better even far from the equatorial plane, but the vertical action $J_z$ in the same approximation is an even better integral, as illustrated in Figure~\ref{fig:staeckel_orbit}. Ultimately, action/angle formalism allows to use a systematic and mathematically well-grounded \textit{nonperturbative} approach based on canonical transformations specified by numerically constructed generating functions \citep{SandersBinney2014,BinneyMcMillan2016} to compute actions to arbitrary accuracy. \citet{SandersEvans2015} compared the quality of self-consistent models constructed using the more accurate action finder with those using the St\"ackel fudge, and concluded that the latter approximation is sufficient in practice.
\end{itemize}

The iterative approach to the construction of self-consistent models with DF specified in terms of actions has been applied by \citet{Binney2014} to a flattened generalization of the Isochrone model. \citet{SandersEvans2015} used a double-power-law DF of \citet{WilliamsEvans2015a} to construct mildly triaxial one-component models.
\citet{Piffl2015, BinneyPiffl2015, ColeBinney2017} extended the approach to a multicomponent model of the Milky Way, using a family of quasi-isothermal DFs (\ref{eq:DFbinney} for the disc population and a double-power-law DF (\ref{eq:DFhalo}) for the dark halo. They recomputed both the halo and the disc density at each iteration; however, to solve the Poisson equation, they approximated the disc density by a separable exponential profile and employed the \textsc{GalPot} approach, hence the potential is only approximately self-consistent and the error cannot be systematically reduced. Moreover, to construct a quasi-isothermal DF, one needs a `seed' potential, which should be reasonably close to the final one. Therefore they employed a two-stage procedure: first the disc density was kept fixed (using the analytic disc profile), while the density and potential of the halo was recomputed iteratively; after a few iterations, the total potential was used to construct the disc DF, and in several more iterations both disc and halo density were recomputed from the DF. Because of the functional form of their quasi-isothermal DF, the final disc density looked rather different from the initial analytic profile, complicating the setup. The need for an intermediate potential for initializing the disc DF also introduced non-trivial dependency between the parameters of the disc and the halo.

%%%%%%%%%%%
\subsection{The present method}  \label{sec:scm_present}

The iterative self-consistent modelling approach in \textsc{Agama} is completely general, although presently restricted to axisymmetric systems due to the availability of action finders. 

A few practical considerations need to be mentioned.
First, the overall potential in a general case is represented by a sum of two approximations -- the azimuthal-harmonic spline expansion for the disc components and the spherical-harmonic spline expansion for the spheroidal components. In the case of an elliptical galaxy model the first one may not be needed, although even elliptical and lenticular galaxies often have significant disky components \citep{RixWhite1990,Emsellem2007}. Each of the two \textit{potential} components, in turn, may correspond to one or several \textit{density} components -- for instance, thin and thick stellar discs for the first one, and stellar bulge and extended dark halo for the second one. In addition, there may be density components that are not specified by their DFs, but are simply `static' contributions to the total potential: for instance, a thin gaseous disc that does not satisfy the collisionless Boltzmann equation, or a `deadweight' softened point mass potential instead of the bulge in the case that our main interest lies outside the central region of the galaxy and we do not need to model the DF of the bulge. The user must explicitly assign each mass component to either of the two potential solvers.

%%%%%%%%%%%%%%
\begin{figure}
\includegraphics{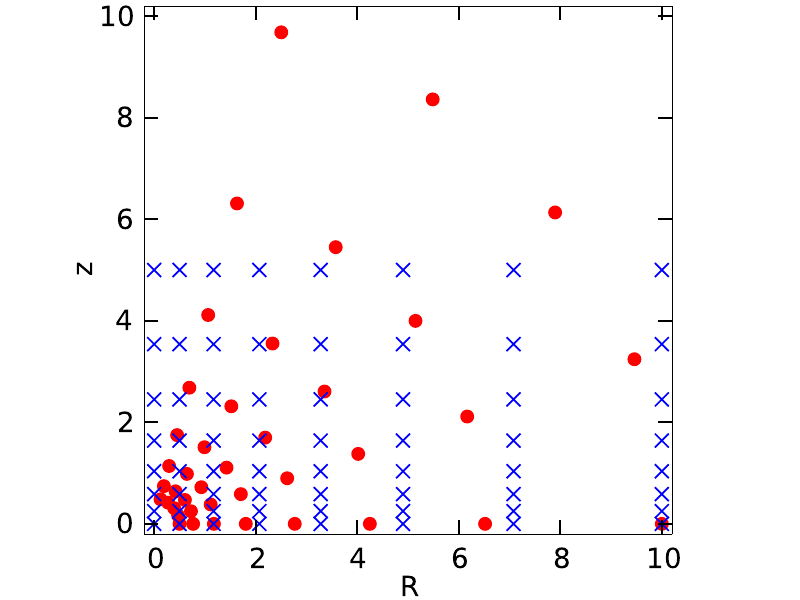}
\caption{Illustration of grids used to represent the density in self-consistent modelling.
Depending on the type of the density profile (spheroidal or disky), we use two different types of grids and correspondingly two classes of potential solvers (spherical-harmonic and azimuthal-harmonic, correspondingly). In the first case, shown by red circles, the density values are collected at the nodes of a logarithmically-spaced grid in $r$ and the nodes of a Gauss--Legendre quadrature rule in $\cos\theta$; in this example we used 8 radial and 5 angular points, corresponding to $l_\mathrm{max}=8$ (only the upper half of the grid is used).
In the second case, shown by blue crosses, the density values are collected at the nodes of a separable grid in $R$ and $z$, which are also spaced non-uniformly, but include a point at the origin.
This arrangement reflects the grids used in the potential interpolation, which is performed in spherical or cylindrical coordinates, respectively. The azimuthal direction ($\phi$) is represented by Fourier harmonics in both cases; however, presently only the $m=0$ azumuthal harmonic is used, because we rely on the axisymmetic St\"ackel fudge in the rest of the modelling framework.
} \label{fig:density_grids}
\end{figure}

Second, in solving the Poisson equation in either spherical-harmonic (\ref{eq:PoissonSphHarm}) or azimuthal-harmonic (\ref{eq:PoissonAziHarm}) form, we need to evaluate the density at many points in space while computing the relevant integrals. However, doing this directly by integrating $f(\boldsymbol{J})$ over velocities at each point (\ref{eq:DensityMoment}) would be extremely inefficient. Instead, we pre-compute the values of density of each component at a moderate number of points on a rectangular 2d grid in $R,z$ or $r,\theta$ planes, and constructing an interpolating spline for fast evaluation of density at any location. This is not quite a trivial step, as the accuracy of potential approximation is determined by the accuracy of density interpolation, and ultimately by the number and location of grid points. For the disc components, 15--30 grid nodes per direction ($R,z$) is enough, provided that the grid covers the spatial region enclosing almost all of the mass of the given component, and the nodes are spaced more densely close to the disc plane where the density is highest (Figure~\ref{fig:density_grids}). For the spheroidal components, we use a uniform grid in log-radius with 20--40 nodes, and nodes of the Gauss--Legendre quadrature rule of order $l_\mathrm{max}+1$ in $\cos\theta$, with the order of multipole expansion $l_\mathrm{max} \simeq 6-8$.

Third, the essence of the method is that the DF is fixed, and gives rise to different density profiles in different potentials. At the same time, expressions for some parametrized DFs (e.g., that of a quasi-isothermal disc, \citealt{BinneyMcMillan2011}) contain the potential implicitly, through the epicyclic frequencies as functions of radius of a circular orbit, which itself is expressed through the actions. This implies that prior to using such a DF, we must compute the dependence of these frequencies on $L_z$ given a plausible initial guess for the total potential.
On the other hand, it is essential to fix this dependence throughout the iterative scheme, so the value of the DF as a function of actions stays constant, even though the epicyclic frequencies in the finally obtained self-consistent potential might be somewhat different from the ones used at the beginning.

In practice, if the goal is to create a model with prescribed density profiles, then one may use quasi-isotropic DFs for spheroidal components and quasi-isothermal DFs for disky components, provided that they were constructed from the desired density profiles in the corresponding total potential. In this case the resulting density profiles are typically only moderately different ($\lesssim 10\%$) from the input ones. However, the DFs are not simply expressible functions, even though they can always be evaluated numerically.
The alternative is to choose some analytic form of DFs, e.g., double-power-law (\ref{eq:DFhalo}) or exponential (\ref{eq:DFexp}), and this completely specifies the resulting model, regardless of the choice of the initial potential: a bad choice may only delay, but not prevent the convergence. Typically, the change in the potential during each subsequent iteration is $\sim 2\times$ smaller, and with a good initial guess, five iterations is enough to be within $1\%$ of the asymptotic solution; even with a poor guess, ten iterations should be sufficient in practice. In terms of wall-clock time, each iteration takes only a few seconds on a 16-core workstation. The present implementation is considerably faster (by a factor of few tens) than the one used in \citet{Binney2014,Piffl2015,BinneyPiffl2015}.

%%%%%%%%%%%
\subsection{Example}  \label{sec:scm_example}

%%%%%%%%%%%%%%%%%
\begin{lstlisting}[
caption={Construction of self-consistent models. \protect\\
We create a single-component model determined by a double-power-law DF (\ref{eq:DFhalo}).
Its density profile resembles an isochrone, and the shape is determined by the coefficients $g_r,g_z,h_r,h_z$ in the linear combination of actions; in this case the axis ratio is $z/x \simeq 0.75$.\protect\\
We start the iterative procedure by providing an initial guess for the density profile (it needs not be close to the final profile, but a good guess speeds up the convergence). Then we perform 5 iterations of recomputation of the density profile followed by reinitialization of the potential and the action finder (all done internally by the \texttt{iterate()} routine). \protect\\
Finally, we create an $N$-body realization of the model by sampling it with $10^5$ particles; this snapshot may be used as initial conditions for an $N$-body simulation.
}, label=code:scm, float=t]
import agama

# the distribution function defining the model
df = agama.DistributionFunction(type='DoublePowerLaw',
    norm=1.7, J0=1, slopeIn=1, slopeOut=5, steepness=2,
    coefJrIn=1.6, coefJzIn=0.8,   # implies coefJphiIn=0.4
    coefJrOut=1.2, coefJzOut=1.2) # -"- coefJphiOut=0.8

# parameters of the radial and angular grid for density
params = dict(rminSph=0.01, rmaxSph=100.0,
    sizeRadialSph=25, lmaxAngularSph=6)

# define the self-consistent model
scm = agama.SelfConsistentModel(**params)

# add a single component to the model and provide
# a reasonable initial guess for its density profile
scm.components = [agama.Component(df=df, disklike=False,
    density = agama.Density(type='Isochrone'), **params)]

# perform several iterations of self-consistent modelling
for i in range(5):
    scm.iterate()

# create and write out an N-body realization of the model
snap = agama.GalaxyModel(scm.potential, df).sample(100000)
agama.writeSnapshot('flattened_isochrone.dat', snap)
\end{lstlisting}
%%%%%%%%%%%%%%%%%

Listing~\ref{code:scm} illustrates the iterative method for constructing self-consistent models with a simple single-component system, resembling the flattened isochrone of \citet{Binney2014}.

We now consider a more interesting example of a three-component axisymmetric bulge--disc--halo galaxy used in section~4.1 of \citet{VasilievAth2015}. In that paper, five different methods were given the task of creating an equilibrium multicomponent model with the given density profile. It has an exponential disc, a flattened S\'ersic bulge with axis ratio $q=0.8$, and a nearly spherical NFW halo with an outer cutoff; the disc is relatively thick and warm (Toomre parameter $Q\gtrsim 1.7$) to minimize the impact of non-axisymmetric instabilities. We created $N$-body realizations of this composite model and checked how close they were to equilibrium by running $N$-body simulations and recording the evolution of radial profiles of three components of the velocity dispersion tensor of disc particles. We found that the Schwarzschild method, implemented in the \textsc{Smile} code \citep{Vasiliev2013}, produced a model that was closest to equilibrium; \textsc{GalactICs} and \textsc{mkgalaxy} models were also acceptable, although the latter one failed to correctly assign the vertical velocity dispersion profile (neglecting contributions from spheroidal components). \textsc{Magalie} \citep{Boily2001} and \textsc{Galic} \citep{YurinSpringel2014} produced models that were significantly out of equilibrium: the former -- due to a simplified treatment of composite potential and a Maxwellian assumption about the velocity distribution, the latter -- because of unrealistic requirement that $\sigma_R=\sigma_\phi$.

%%%%%%%%%%%%%%
\begin{figure}
\includegraphics{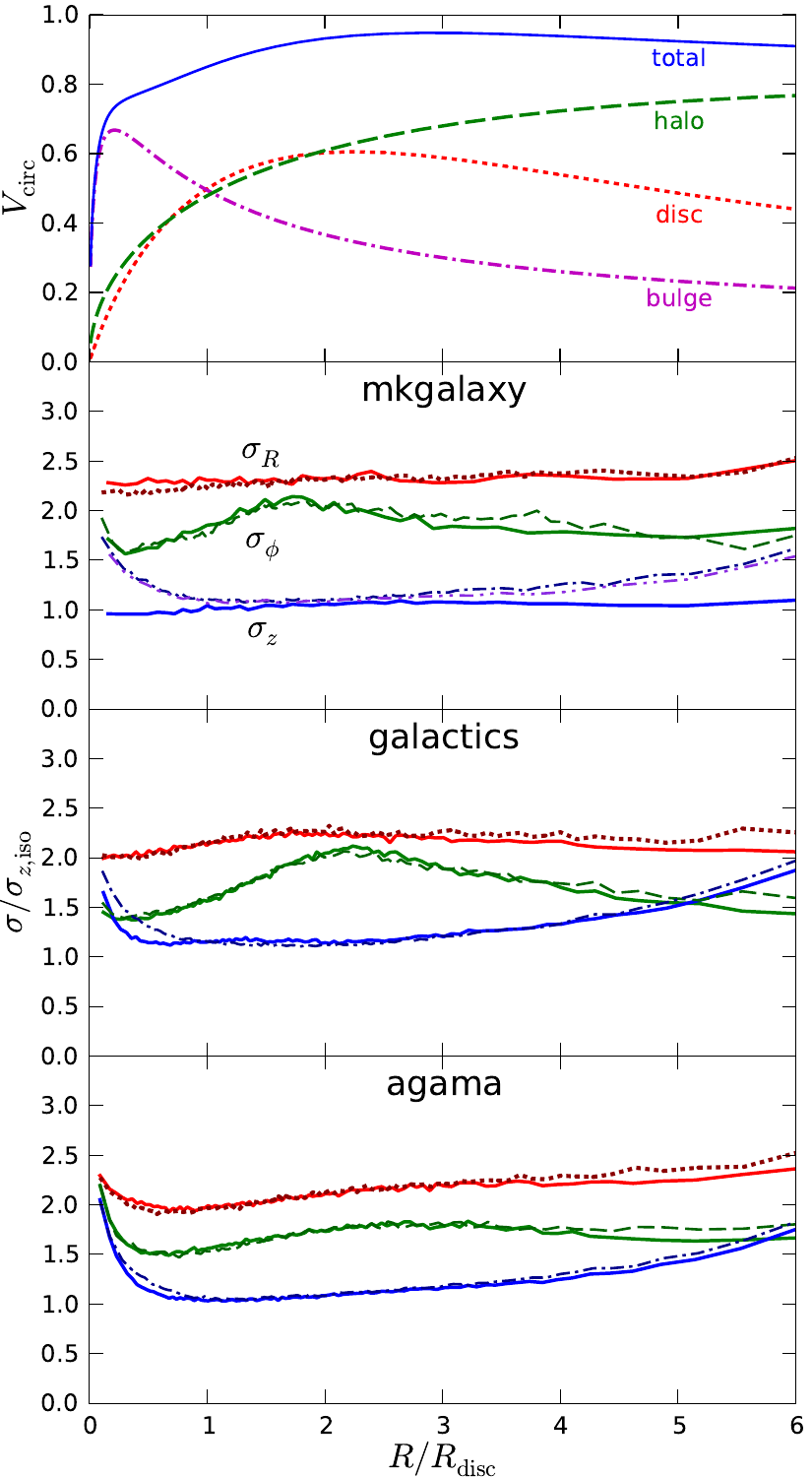}
\caption{Self-consistent disc--bulge--halo models constructed by three different methods: \textsc{mkgalaxy} \citep{McMillanDehnen2007}, \textsc{GalactICs} \citep{Widrow2008}, and \textsc{Agama} (this paper).
Top panel shows the intended rotation curve, which is reproduced with minor deviations by all codes.
Other panels show the radial profiles of three components of velocity dispersion tensor of the disc component, normalized by the value of vertical velocity dispersion that would correspond to an isolated thin disc: $\sigma_\mathrm{z,iso} \equiv \sqrt{G\, M_\mathrm{disc}\,h / R_\mathrm{disc}^2}\:
\exp\big(-\frac{1}{2} R/R_\mathrm{disc}\big)$, where $M_\mathrm{disc}=1$, $R_\mathrm{disc}=1$, $h=1/16$ are the mass, scale radius and scale height of the disc. Solid lines correspond to the initial models, and dashed lines -- to the models evolved for 100 time units. Dash-double-dotted line in the second panel shows the vertical velocity dispersion profile for a model evolved in a fixed (initial) potential, which is noticeably higher than the initial one; for other methods and velocity components, the fixed-potential evolution leaves the profiles essentially unchanged and they are not shown to avoid crowding. Except for this one case, the evolution of live models leads only to rather moderate structural changes, mainly in the outer parts which develop some degree of spiral instability. Figure~2 in \citet{VasilievAth2015} shows three other methods in the same comparison project (note that $\sigma_R$ and $\sigma_\phi$ are erroneously swapped in that figure).
} \label{fig:scm}
\end{figure}
%%%%%%%%%%%%

We now add another method to this test suite, using the same density profile to initialize quasi-isotropic DFs of the spheroidal components (bulge and halo) and a quasi-isothermal DF of the disc, and performing 5 iterations of the self-consistent modelling procedure. The resulting density profiles are not exactly the same as the initial ones, but the difference is $\lesssim 10\%$. We then created an $N$-body realization of the model and evolved it for 100 time units with the fast-multipole code \textsc{gyrfalcON} \citep{Dehnen2000}, as in the previous study. To minimize the impact of relaxation on the disc structure, we increased the number of particles five-fold compared to the previous paper, using 0.2, 0.8 and 4 million particles for the bulge, disc and halo components, respectively. 

Figure~\ref{fig:scm} shows the initial and evolved velocity dispersion profiles of the disc component, comparing \textsc{Agama} with two other similar methods (\textsc{GalactICs} and \textsc{mkgalaxy}), which were also re-run with a higher number of particles. The initial profiles are not exactly the same due to the differences in the methods, but they are close enough for a meaningful comparison. All three models displayed some evolution caused primarily by the development of a rather moderate spiral instability in the outer parts of the disc. In general, though, the changes in velocity dispersion profiles and other dynamical characteristics were minor, indicating that the initial conditions were in a good equilibrium. To verify this, we computed the orbits of disc particles for the same time interval in a fixed potential of all three components, using the two potential approximations constructed from the particles of the initial $N$-body snapshot for each method. The final states of these fixed-potential models were very close to the initial ones (we do not plot the respective curves because they very nearly coincide), except for the vertical velocity dispersion in \textsc{mkgalaxy} (shown as a dash-double-dotted line), which, as mentioned above, was underestimated in the initial conditions.

This example illustrates that the DF-based self-consistent method in \textsc{Agama} produces models that are at least as good as the other commonly used methods; however, it offers much greater flexibility in choosing the model parameters, number and type of components, etc. It is also competitive in terms of computational effort, taking about one minute of wall-clock time on a 16-core workstation.

%%%%%%%%%%%%%%%%%%%%
\section{Discussion}  \label{sec:discussion}

We presented a software framework for stellar dynamics and galaxy modelling. It provides general-purpose methods for constructing smooth potentials from arbitrary density distributions or from $N$-body snapshots, conversion between position/velocity and action/angle variables (presently only for spherical or oblate axisymmetric potentials), distribution functions and their moments, iterative construction of multicomponent self-consistent models. 

Methods that are new or considerably improved compared to previous works include:
\begin{itemize}
\item Efficient and versatile general-purpose potential representations in terms of spherical-harmonic and azimuthal-harmonic expansion. They can be used to accurately compute the potential of almost any density profile provided in a functional form, as well as construct a smooth 
approximation to the potential of an $N$-body system. The first of these approaches is a more flexible alternative to the familiar basis-set expansion technique, while the second has been little used so far in the astrophysical community, but is highly suitable for disky galaxies, especially non-axisymmetric.
\item A novel implementation of the St\"ackel fudge approach for computing approximate action/angle variables in arbitrary axisymmetric potentials, which is more accurate and efficient than other existing codes.
\item Several types of action-based DFs, including improvements and generalizations of previously proposed ones, and a new class of disc DF formulated entirely in terms of actions.
\item Various tools for working with DFs: optimized computation of DF moments and 1d marginalized velocity distributions, general routines for sampling from a multidimensional probability distribution and for non-parametric penalized density estimates from discrete samples.
\item Framework for iterative construction of DF-based multi-component self-consistent models, extending previous similar approaches and improving both the performance and flexibility.
\end{itemize}

The \textsc{Agama} library is written in C++, provides Python and Fortran interfaces and bindings to several other stellar-dynamical packages. Considerable effort is invested into computational efficiency and mathematical robustness. Most time-consuming operations are internally parallelized using \texttt{OpenMP}, both when using the library natively in C++ programs or as a Python extension module.

%%%%%%%%%%%
\subsection{Comparison to other similar projects}

There are several other software projects with broadly similar scope.

The most well-known is \textsc{Galpy} \citep{Bovy2015} -- a Python package for galactic dynamics, offering a large collection of gravitational potentials, orbit integration routines, methods for conversion between position/velocity and action/angle coordinates, action-based distribution functions, coordinate conversion, and plotting facilities. As evident from this list, there is a lot of overlap with the features provided by \textsc{Agama}. \textsc{Galpy} has a larger variety of built-in potentials, however, \textsc{Agama} offers two general-purpose potential solvers that can suit almost any imaginable need, and can be constructed from arbitrary user-provided density profiles as well as from $N$-body snapshots. Both libraries have several methods for action computation: isochrone, arbitrary spherical potentials, St\"ackel fudge for axisymmetric potentials. \textsc{Galpy} additionally offers adiabatic approximation (which is generally inferior compared to the St\"ackel fudge) and a method based on canonical transformation using numerically integrated orbits, which is more accurate but far more expensive; on the other hand, the most practical St\"ackel approximation in \textsc{Agama} is more accurate and substantially faster. \textsc{Galpy} is primarily written in Python, but for greater efficiency parts of the code (some potentials and action finders) are also re-written in C, which leads to a rather complicated internal structure and code duplication. By contrast, all computationally intensive parts of \textsc{Agama} are written in C++; the Python interface exposes most of the underlying functionality without sacrificing the performance, but offering greater flexibility (for instance, a user-provided Python function can be used as a density profile for the two general-purpose potential solvers). Perhaps the best approach is to combine the advantages of both packages -- for instance, any \textsc{Agama} potential can be used as a regular \textsc{Galpy} potential (the library provides a simple wrapper), its action finders can also serve as more efficient substitutes to those from \textsc{Galpy}, while the latter provides nice visualisation features.

Another package with similar functionality is \textsc{Gala} \citep{PriceWhelan2017}, written in Cython (a more performance-oriented superset of Python that is compiled just like C/C++ code). It also offers a collection of gravitational potentials (most of them have counterparts in \textsc{Agama}), orbit integration routines, and an orbit-based action transformation method of \citet{SandersBinney2014}.

\textsc{Tact} \citep{SandersBinney2015a} is a C++ library with the broadest collection of action transformation methods; those missing from \textsc{Agama}, such as the triaxial St\"ackel fudge or the generating function method, can be easily coupled to it thanks to the shared language and similar conventions.

%%%%%%%%%%%
\subsection{Possible usage scenarios}

\begin{itemize}
\item Approximate the potential of a snapshot from an $N$-body simulation with a smooth non-parametric potential expansion, which faithfully represents global features while discarding small-scale noise. This drastically reduces the amount of information needed to represent the `frozen' potential of the given system, which can then be used to integrate and classify orbits of particles. This task traditionally involved using the same Poisson solver as in the simulation itself (e.g., a tree code), as in \citet{Valluri2010, Valluri2016, Portail2015}, or approximating the potential using a basis-set expansion \citep[e.g.,][]{Hoffman2010, Bryan2012, Roettgers2014}, or fitting it by some combination of analytic models \citep[e.g.,][]{Muzzio2005, MachadoManos2016, Maffione2018}.
The $N$-body potential solvers are both expensive and noisy, and analytic models or conventional basis sets do not offer enough flexibility and could lead to significant biases \citep[e.g.,][]{CarpinteroWachlin2006, Kalapotharakos2008}. Our spline-interpolated spherical- and azimuthal-harmonic expansions provide an optimal balance between efficiency and accuracy of potential approximation; the first of these is used in \citet{Zhu2017}, le Bret et al.(in prep.).
\item These potential solvers can also be applied to smooth density models, in which the alternative approaches for solving the Poisson equation may be expensive, such as a triaxial bar or spiral arms \citep[e.g.,][]{Pichardo2003,Antoja2011,Fragkoudi2015}. They are also used for orbit computation in the \citet{Schwarzschild1979} orbit-superposition method, implemented in the \textsc{Smile} code \citep{Vasiliev2013}, which is included in the library.
\item Instead of classifying numerically-integrated orbits, one may compute particle actions and use clustering algorithms to detect substructure such as clumps or streams \citep[e.g.,][]{Sanderson2015, Helmi2017}.
\item DF-based models provide a compact representation of the full 6d phase space of $N$-body models, similar to what smooth potentials do for the 3d density profile. At present, \textsc{Agama} provides either non-parametric spherical isotropic DFs (one-dimensional), or a choice from several families of parametric three-integral models.
\item Spherical isotropic DFs, while highly idealized, nevertheless can serve as a good approximation to some collisional systems, such as globular clusters. They can be used to compute classical two-body relaxation coefficients, which have been shown to describe quite well the actual evolution of corresponding $N$-body systems \citep[e.g.,][]{Vasiliev2015}. Using the tools from \textsc{Agama}, \citet{Beraldo2017} demonstrated the agreement between entropy evolution measured from $N$-body simulations and predicted by the smooth DF extracted from the simulations.
\item DF-based models are well suited for describing observed stellar systems. A prime example is our Galaxy, where such models have been fitted to a vast collection of observational data (e.g., the Besan\c con model, \citealt{Robin2003}). 
Much effort has gone into determination of suitable DFs describing the kinematics of stars in the solar neighbourhood \citep[e.g.,][]{Binney2010, Sharma2014, SandersBinney2015b, Posti2018} or in the galactic halo \citep[e.g.,][]{WilliamsEvans2015b, DasBinney2016, Das2016}, assuming that the gravitational potential is known and focusing on velocity anisotropy and rotation, or relations between age, chemical composition and kinematics of different stellar populations.
\item One may instead consider the stars as kinematic tracers of the potential, and use their motion to constrain the total distribution of matter (both visible and dark). This is often performed using Jeans equations, but DF-based models have an advantage of always providing a physically valid solution (which is not guaranteed in the Jeans approach). \citet{McMillanBinney2013, Ting2013, Piffl2014, Trick2016} constructed action-based DF models of the Milky Way, varying the parameters of both the DF and the potential when fitting them to the observations. The DF approach has also been applied to infer the potential of nearby dwarf spheroidal galaxies from radial velocities of individual stars \citep[e.g.][]{Wilkinson2002, WalkerPenarrubia2011, AmoriscoEvans2011}, and \citet{Pascale2018} pioneered the use of action-based self-consistent models for one of these systems, although only in a spherical geometry. In this approach the parameters of DF and the potential are varied independently, and the contribution of stars to the total potential is typically ignored as negligible.
\item Ultimately, the complete galaxy model must combine the potential and the DFs of all components (including dark matter) in a dynamically self-consistent way. \citet{Piffl2015, BinneyPiffl2015, ColeBinney2017} used the iterative method with action-based DFs to construct such models of Milky Way. \textsc{Agama} brings this method to a new level by substantially improving the performance and abandoning the ad hoc approximation of separable disc density used in computing its potential, in favour of more general potential solvers. These models can be fitted to a variety of observational constraints and hence present the most general way of self-consistently describing the Galaxy in terms of just a handful of parameters (Binney \& Vasiliev, in prep.).
\item The same method may be used to construct an equilibrium galaxy model with known (or assumed) DFs and then use it to generate an $N$-body snapshot. In this context it is similar to the \textsc{GalactICs} code \citep{KuijkenDubinski1995, Widrow2008, Taranu2017} but more flexible and potentially more accurate thanks to the general-purpose potential solvers. The formulation of the DF in terms of actions, as opposed to any other integrals of motion, is also beneficial especially for construction of multicomponent models, which can be done simply by adding together several DFs and readjusting the total potential without modifying the explicit expressions for DFs.
\item Axisymmetric models are only an approximation for real disc galaxies (such as Milky Way), but they could serve as a suitable starting point for a more complicated analysis. In this aspect, an explicitly known three-integral DF expressed in terms of actions opens up many possibilities, e.g., using perturbation theory to study deviations from axisymmetric equilibrium state \citep[e.g.,][]{Monari2016, Binney2018}, normal modes and instabilities of self-gravitating discs \citep[e.g.,][]{Kalnajs1977, Polyachenko2005, JalaliHunter2005}, or the impact of resonances on the relaxation of dynamically cold systems \citep[e.g.,][]{Fouvry2015}.
\item DF-based models are a valuable tool in generating mock catalogues for the given survey parameters (e.g., the \textsc{Galaxia} code, \citealt{Sharma2011}). These mock datasets may be used to test the performance of the chosen modelling approach (\textsc{Agama} was used in this context by \citealt{Zhu2016} to validate their discrete Jeans models). The method for sampling from a multidimensional probability distribution (the product of a DF and a selection function of the survey) provided in \textsc{Agama} is completely general and quite efficient, potentially superseding earlier schemes tuned for a particular form of DF.
\end{itemize}

%%%%%%%%%%%
\subsection{Advantabes and limitations}

The approach to dynamical modelling based on a DF has several advantages both in theoretical and observational applications. A DF is a smooth representation of a stellar system and can be treated as a probability distribution in various contexts. One is the creation of equilibrium $N$-body models with arbitrary large number of particles, something that cannot be easily done with the \citet{Schwarzschild1979} or made-to-measure (M2M, \citealt{SyerTremaine1996}) models, which have a finite number of discrete elements. The other comes into play when comparing models to data using a likelihood approach. In general, there seems to be no practical way of doing this if both the model and the data points are discrete samples of some underlying probability distribution, without smoothing or binning either of them (see \citealt{Saha1998}, \citealt{Chaname2008} and \citealt{deLorenzi2008} in the context of $N$-body, Schwarzschild and M2M models, respectively). It has also been argued that the discrete nature of Schwarzschild-type models presents challenges in likelihood-based inference of high-quality data such as that for stars in the Milky Way \citep{McMillanBinney2013}, which are avoided in models with a smooth DF. Finally, it is easy to compare different models specified in terms of a DF with a known functional form, while this hardly can be done if the model consists of individual particles or orbits.

Of course, this approach is not without limitations. It relies on the fundamental assumption of integrability of motion in the given potential. Clearly, this assumption is violated in triaxial systems, such as elliptical or barred disc galaxies, but even in purely axisymmetric systems the phase space may contain multiple orbit families separated by chaotic layers. Nevertheless, the formalism of action/angle variables may be meaningfully used even in this case \citep[e.g.,][]{Kaasalainen1995a,Kaasalainen1995b,Binney2016}. Still, even ignoring the complications arising from non-integrability, one may argue that if the method used to compute actions behaves regularly in the presence of resonant and chaotic orbits, then the description in terms of a smooth DF remains approximately valid for the task of creating a (nearly) self-consistent model \citep{Binney2018}. %One should note that similar concerns apply to Schwarzschild models, in which the existence of chaotic orbits complicates the construction and interpretation of models \citep{Pfenniger1984,MerrittFridman1996}, but does not by itself imply that they should be treated differently from the regular orbits \citep{VasilievAth2012}.
Other limitations of the present implementation, specific to action computation, are its restriction to axisymmetric systems and the use of the St\"ackel fudge. They are not fundamental and may be lifted in the future; the work of \citet{SandersEvans2015} demonstrates both the extension of iterative DF-based method to non-rotating triaxial models of elliptical galaxies and the use of a more accurate action determination algorithm, which was found not to change the results appreciably.

\vspace{5mm}
To summarize, the \textsc{Agama} framework complements (and in some aspects supersedes) the existing software libraries for galaxy modelling, and can be used in many theoretical and observationally-motivated applications. As with most scientific software, it is being continuously developed; we hope that releasing it publicly will be beneficial for the community, and welcome the feedback and suggestions for further extension and improvement.

I am grateful to L.~Beraldo e Silva, J.~Binney, D.~Cole, P.~Das, V.~Debattista, W.~Dehnen, K.~Hattori, E.~Polyachenko, L.~Posti, J.~Sanders, M.~Valluri, the anonymous referee, and other colleagues for valuable comments on the paper and continuous feedback on the code.
This work was supported by the European Research council under the 7th Framework programme  (grant No.\ 321067).

\end{document}